\documentclass[a4paper,aps,prd,twocolumn,eqsecnum,amsmath,amssymb,nofootinbib,superscriptaddress]{revtex4-2}
\usepackage{dcolumn}
\usepackage{bm}
\usepackage{graphics}
\usepackage{graphicx}
\graphicspath{{figures/}}
\usepackage{epsfig}
\usepackage{booktabs,multirow}
\usepackage{hhline}
\usepackage{slashed}
\usepackage{rccol}
\usepackage{mathrsfs}
\usepackage{amsmath, amssymb}
\usepackage[dvipsnames]{xcolor,colortbl}
\usepackage{subfigure}
\newcommand{\orcid}[1]{\href{https://orcid.org/#1}{\includegraphics[width=10pt]{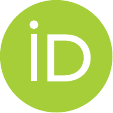}}}

\usepackage{xcolor}\colorlet{linkequation}{Blue}
\usepackage[colorlinks=true, 
linkcolor=Red,   
filecolor=Red,   
citecolor=Green, 
urlcolor=Blue,
hyperfootnotes=true]{hyperref}

\newcommand*{\SavedEqref}{}
\let\SavedEqref\eqref
\renewcommand*{\eqref}[1]{%
	\begingroup
	\hypersetup{
		linkcolor=linkequation,
		linkbordercolor=linkequation,
	}%
	\SavedEqref{#1}%
	\endgroup
}

\begin{document}
	\title{Constraints on active-sterile neutrino transition magnetic moments from low-energy electronic recoils at direct detection experiments}

	\author{M. F.~Mustamin \orcid{0000-0003-3996-4651}}
	\email{mfmustamin@ktu.edu.tr}
	\affiliation{Department of Physics, Karadeniz Technical University, Trabzon 61080, Türkiye}
	\affiliation{Graduate School of Natural and Applied Science, Karadeniz Technical University, Trabzon 61080, Türkiye}
	
	\author{M.~Demirci  \orcid{0000-0003-2504-6251}}
	\email{mehmetdemirci@ktu.edu.tr}%
	\affiliation{Department of Physics,
		Karadeniz Technical University, Trabzon 61080, Türkiye}
	
	\date{\today}

\begin{abstract}
	Sterile neutrinos can potentially be produced through neutrino transition magnetic moments in neutrino-electron scattering. In this work, we investigate this dipole portal by analyzing low-energy electronic recoil data from the PandaX-4T (Run0 and Run1) and XENONnT experiments. We focus on the up-scattering of solar neutrinos into sterile states via the transition magnetic moment. By performing a comprehensive analysis, we derive robust exclusion limits on the neutrino flavor-independent and flavor-dependent transition magnetic moments for fixed sterile neutrino masses as well as on the mass-coupling parameter space. We demonstrate that, since they can detect solar neutrino-induced processes, direct detection experiments offer a unique framework for studying all possible neutrino flavors. The obtained limits extend the sensitivity to previously unexplored regions of the parameter space.
\end{abstract}


\maketitle

\section{Introduction}

Direct detection (DD) experiments, which are intended to search for dark matter (DM), are undergoing rapid development. These detectors are typically located deep underground to mitigate background from cosmic rays and cosmogenic radiation. Their primary goal is to measure electronic or nuclear recoil signals resulting from potential interactions between dark matter particles and detector target materials. Moreover, as neutrinos constitute an irreducible background in these experiments, DD facilities offer a valuable platform to probe neutrino properties. In particular, they enable the study of solar neutrino interactions, allowing precision measurements of the Standard Model (SM) weak mixing angle at the lowest accessible energies, and provide sensitivity to potential contributions from new neutrino interactions beyond the SM (BSM).
The discovery of neutrino oscillations~\cite{Super-Kamiokande:1998kpq, SNO:2001kpb, SNO:2002tuh} provides compelling evidence that neutrinos possess nonzero masses. This observation underscores the necessity to extend the SM, as the minimal framework predicts massless neutrinos due to the absence of right-handed singlets and the consequent conservation of total lepton number.

In many BSM frameworks, the mechanisms that generate neutrino masses invoke the existence of additional heavy neutral leptons, commonly referred to as sterile neutrinos~\cite{Pontecorvo:1967fh, Kusenko:2009up, Dasgupta:2021ies}.
The sterile neutrino hypothesis is motivated by its potential to resolve several anomalies observed in short-baseline oscillation and reactor facilities, including MiniBooNE \cite{MiniBooNE:2010idf}, MicroBooNE \cite{Arguelles:2021meu}, and LSND \cite{LSND:1997vun}. The proposed particle, capable of explaining these anomalies, would typically possess mass in the eV scale, with a possible connection to nucleosynthesis processes in core-collapse supernovae \cite{McLaughlin:1999pd, Xiong:2019nvw}. Furthermore, sterile neutrinos in the higher mass regime have been proposed as viable dark matter candidates \cite{Dodelson:1993je}. Beyond these contexts, the sterile neutrino framework has been extensively studied in various theoretical scenarios, including its role in inducing effective neutrino magnetic moments \cite{Balantekin2014}, connections to CP violation \cite{Balaji:2020oig}, extra dimensions \cite{Khan:2022bcl}, effective field theory \cite{Duarte:2023tdw}, and its influence on the early universe's thermal history and evolution \cite{Mirizzi:2012we}.

One possible process for producing sterile neutrinos is through the up-scattering \cite{Domokos:1996cn, Gninenko:1998nn} of active neutrinos off electrons via transition magnetic moments \cite{McKeen2010}. This process is conceptually related to the Primakoff up-scattering mechanism, originally proposed for studying photoproduction of neutral mesons in a nuclear electric field \cite{Primakoff:1951iae}. Investigating this phenomenon via electronic recoil signals in DD experiments is of particular interest, as it offers a novel avenue to constrain active-sterile neutrino transition magnetic moments and related parameters.
This transition magnetic moment has been broadly studied using data from a wide range of experiments. Some of them include accelerator-based neutrino sources \cite{Magill:2018jla, Blanco2020, Schwetz2021}, neutral-current $\nu_\mu$–nucleus scattering experiments \cite{Gninenko2011}, spallation neutron sources \cite{DeRomeri:2022twg}, nuclear reactors \cite{DeRomeri:2025csu, AristizabalSierra:2022axl, Bolton2022}, forward neutrino detection facilities \cite{Ismail:2021dyp}, high-energy particle colliders \cite{Antusch:2016ejd, Brdar:2025iua}, and atmospheric neutrino observations \cite{Coloma:2017ppo, Plestid:2020vqf, Gustafson:2022rsz, Atkinson:2021rnp}. Additionally, several studies investigated this scenario using previous DD results \cite{Miranda2021, Li2022}, as well as utilizing cosmological neutrino sources \cite{Brdar:2020quo}, neutrino telescope observations \cite{Huang:2022pce}, and the diffuse supernova neutrino background \cite{Balantekin:2023jlg}.

Prompted by the aforementioned motivations, we probe the contribution of active-sterile neutrino transition magnetic moments to the elastic neutrino–electron scattering (E$\nu$ES) framework, utilizing recent low-energy electronic recoil signals from DD experiments. As one of the irreducible backgrounds in DD experiments, E$\nu$ES events induced by solar neutrinos provide a natural framework to probe active-sterile neutrino transition magnetic moments. 
Particularly, we focus on the PandaX-4T \cite{PandaX:2024cic} and XENONnT \cite{XENON:2022ltv} recent datasets, deriving new limits on the transition magnetic moment and further tightening available results in the literature. We have applied these datasets in our previous studies \cite{Demirci:2025qdp, Demirci:2025nep} to explore other new physics scenarios.
Moreover, given that solar neutrino oscillations allow the incoming flux to exhibit a flavor composition at Earth, we perform both flavor-independent and flavor-dependent analyses. 
Furthermore, we discuss the derived limits with the existing constraints obtained from various facilities. These include data from CHARM-II \cite{CHARM:1983ayi}, NOMAD \cite{NOMAD:1997pcg}, BOREXINO \cite{Borexino:2007kvk, Borexino:2008fkj}, TEXONO \cite{TEXONO:2009knm}, IceCube \cite{IceCube:2015vkp}, Dresden-II \cite{Colaresi:2022obx}, and CONUS+ \cite{Ackermann:2025obx}. We also provide limits from sterile neutrino decay searches using solar \cite{BOREXINO:2018ohr} and atmospheric \cite{Super-Kamiokande:2017yvm} neutrinos, and recent CE$\nu$NS-induced solar neutrino measurements at PandaX-4T \cite{PandaX:2024muv} and XENONnT \cite{XENON:2024ijk}, as well as other astrophysical and cosmological bounds.

The outline of this work is the following. We briefly explain in Sec.~\ref{sec:eves} the E$\nu$ES process in the SM and in the up-scattering process through the active-sterile neutrino transition magnetic moment. We then provide the details for data analysis in Sec.~\ref{section:data}. 
In Sec.~\ref{sec:results}, we discuss the predicted event rates and our derived limits of the parameters. We finally conclude our work in Sec.~\ref{sec:conc}.

\section{Theoretical Framework}\label{sec:eves}
\subsection{E$\nu$ES in the SM}
The process of E$\nu$ES has a nature of being extremely directional. The scattered electrons have very small angles relative to the incoming neutrinos' direction \cite{Formaggio:2012cpf}. This property has been utilized in many neutrino experiments, particularly for solar neutrino detections at Super-Kamiokande \cite{Kamiokande-II:1989hkh, Super-Kamiokande:2001ljr}, SNO \cite{SNO:2003bmh, SNO:2008gqy}, and BOREXINO \cite{Borexino:2007kvk, Borexino:2008fkj}.
Furthermore, the coming solar neutrinos can induce E$\nu$ES events in DD experiments of DM as they collide with the target material in a detector. This process, together with coherent neutrino-nucleus scattering, is one of the neutrino background components in the DD experiments. 
Therefore, signals of neutrino-electron scatterings have a degeneracy with the DM-electron scatterings.

\begin{figure}[htb]
	\centering
	\includegraphics[scale=0.58]{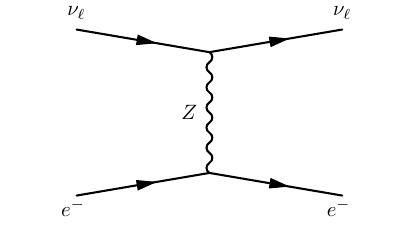}
	\includegraphics[scale=0.58]{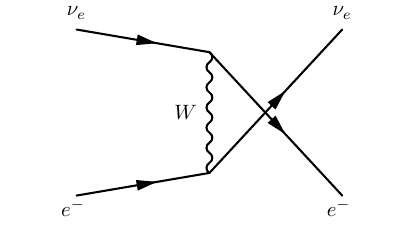}
	\\
	(a) \hspace{36mm} (b)
	\\
	\caption{Representative diagrams for E$\nu$ES in the SM via (a) NC and (b) CC channels.
    }
	\label{fig:nue_diag}
\end{figure}
The E$\nu$ES processes are purely leptonic interactions within the SM. We present the tree-level Feynman diagrams of the scatterings in Fig.~\ref{fig:nue_diag}. The index $\ell$ runs for $e,\mu$, or $\tau$ flavor of neutrino. The coming neutrinos $\nu_\ell$ scatter off electrons $e^-$ through charged-current (CC) or neutral-current (NC) channels, depending on the neutrino flavor. 
The neutral gauge boson is represented by $Z$, while the charged gauge boson is $W$. The differential cross-section of the SM process is given by
	\begin{align} 
		\begin{split}
			\left[\frac{d\sigma_{\nu_\ell}}{dT_{e}}\right]_{\mathrm{SM}} = \frac{G_F^2 m_e}{2\pi} &\Bigg[ (g_V+g_A)^2 + (g_V-g_A)^2 \left(1-\frac{T_{e}}{E_\nu}\right)^2  \\ &-(g_V^{2}-g_A^{2}) \frac{m_e T_{e}}{E_\nu^2} \Bigg],
		\end{split}
		\label{eq:sm_nue}
	\end{align}
where $T_{e}$ is the electron recoil energy, $m_e$ is the electron mass, $E_\nu$ is the energy of the incoming neutrino, and $G_F$ is the Fermi coupling constant 
\cite{ParticleDataGroup:2024cfk}. 
%
The vector and axial-vector couplings to specific neutrino flavors are
\begin{align}
	g_V = -\frac{1}{2}+2s_W^2 + \delta_{\ell e}  \quad \text{and} \quad
	g_A = -\frac{1}{2}+\delta_{\ell e},
\end{align}
respectively. We adopt the abbreviation $s_W=\sin\theta_W$, where $\theta_W$ denotes the weak mixing angle.  
These couplings depend on the incoming neutrino's flavor. 
The Kronecker delta symbol $\delta_{\ell e}$ stands for the NC and CC nature of the process. 
Taking radiative corrections into account, we use flavor-dependent couplings $g_V^{\nu_e e}=0.9524, g_A^{\nu_e e}=0.4938$, $g_V^{\nu_\mu e}=-0.0394$, $g_A^{\nu_\mu e}=-0.5062, g_V^{\nu_\tau e}=-0.0350$, and $g_A^{\nu_\tau e}=-0.5062$ \cite{Erler:2013, Tomalak:2019ibg, AtzoriCorona:2022jeb}.
We emphasize that for $\nu_e$ both NC and CC contribute to the calculation, while for the $\nu_\mu$ and $\nu_\tau$ only the NC contributes.

\subsection{Sterile neutrino dipole portal}
The active-sterile neutrino transition magnetic moments could be investigated by electromagnetically up-scattered neutrino beams on electrons. The effective operator at low-energy of such interactions can be written as \cite{Bolton2022}
\begin{eqnarray} \label{eq:lag}
	\mathcal{L}_\text{int} \supset \frac{\mu_{\nu_{\ell 4}}}{2} \bar{\nu}_{\ell L} \sigma^{\mu\nu} P_R \nu_4 F_{\mu\nu} + \text{h.c.}.
\end{eqnarray}
The coupling of the transition magnetic moment is represented by $\mu_{\nu_{\ell 4}}$, while $\nu_4$ denotes the sterile neutrino, $\nu_{\ell L}$ the SM neutrino, and $F_{\mu\nu}$ the electromagnetic field tensor. 
It should be emphasized that this form is suitable at low energies only, typically less than the electroweak (EW) energy. The E$\nu$ES process considered here involves an incoming neutrino with low energy, and hence the formulation remains applicable.

\begin{figure}[htb]
	\centering
	\includegraphics[scale=0.75]{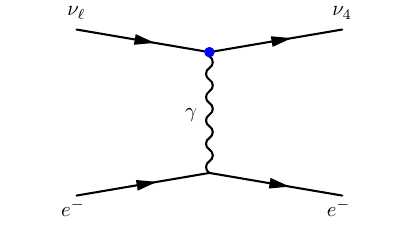}
	\caption{Representative diagram for the up-scattering of $\nu_\ell + e^- \rightarrow \nu_4 + e^-$. The dotted vertex represents the neutrino dipole portal of the active-sterile neutrino transition magnetic moment. 
    }
	\label{fig:upscattering}
\end{figure}
We show in Fig. \ref{fig:upscattering} the representative diagram of the up-scattering process of E$\nu$ES through $\mu_{\nu_{\ell 4}}$.
The initial active neutrino $\nu_\ell$, through a sterile neutrino dipole portal, exchanges a photon with the target electron and up-scatters to the final sterile neutrino $\nu_4$. We can write the process's amplitude as 
\begin{eqnarray} \label{eq:amp}
	i\mathcal{M} = \mu_{\nu_{\ell 4}} \left(\bar{u}_{\nu_4} \sigma^{\mu\nu} P_L q_\nu u_{\nu_\ell}\right) \left(\frac{-ig_{\mu\lambda}}{q^2}\right) j_{e}^\lambda,
\end{eqnarray}
with the leptonic current $j_e^\lambda = -ie (\bar{u}_e \gamma^\lambda u_e)$.
The differential cross-section of the process is
\begin{equation}
\begin{split}
    \biggl[\frac{d\sigma (E_\nu, T_{e})}{dT_{e}} \biggr] = &\frac{\pi \alpha_{\mathrm{EM}}^2}{m_e^2} \left| \frac{\mu_{\nu_{\ell 4}}}{\mu_B} \right|^2  \Bigg[\frac{1}{T_{e}} - \frac{1}{E_\nu}  \\ &- \frac{m_4^2}{2T_{e}E_\nu m_e } \left(1-\frac{T_{e}}{2E_\nu}+\frac{m_e}{2E_\nu}\right) \\ &- \frac{m_4^4}{8 m_e T_{e}^2 E_\nu^2} \left(1-\frac{T_{e}}{m_e} \right) \Bigg],
\end{split} \label{eq:sterile_cs}
\end{equation}
where the factor of $\pi \alpha_{\mathrm{EM}}/m_e^2$ arises from normalizing the magnetic moments in terms of the Bohr magneton, $\mu_B$.
Additionally, there is a kinematic constraint of the sterile neutrino mass
\begin{eqnarray}\label{eq:m4const}
	m_4^2 \leq 2m_e T_{e} \left(E_\nu\sqrt{\frac{2}{m_e T_{e}}} -1 \right).
\end{eqnarray}
The contribution from an active-sterile neutrino transition magnetic moment is added coherently to the E$\nu$ES in the SM, leading to potential enhancements in the predicted event at low-energy electronic recoil signals.

It is worth noting that the terms dependent on the sterile neutrino mass $m_4$ in Eq. \eqref{eq:sterile_cs} significantly alter the spectral shape compared to the standard active-active scattering \cite{Vogel:1989iv}. While the massless limit ($m_4\rightarrow 0$) exhibits a characteristic $1/T_e$	enhancement at low recoil energies, the mass-dependent terms introduce a kinematic suppression near the threshold. This distinction allows for a specific characterization of the sterile neutrino signal, differentiating it from standard electromagnetic neutrino interactions.

\section{Data Analysis Details}
\label{section:data}
\subsection{Event Rate}
We express the differential event rate per unit target mass as
\begin{align}
\left[\frac{dR}{dT_{e}}\right]^i = Z_{\text{eff}}(T_e) \int_{E_{\nu}^{\text{min}}}^{E_{\nu,i}^{\text{max}}} dE_\nu \frac{d\Phi^i}{dE_\nu} \left[\frac{d\sigma}{dT_{e}}\right],
\end{align}
where $d\Phi^i/dE_\nu$ is the differential neutrino flux for the $i^{th}$ solar neutrino component, and $Z_{\mathrm{eff}}(T_e)$ denotes the effective atomic number accounting for the number of available electrons per recoil energy bin in the detector. The differential cross section $d\sigma/dT_{e}$ includes contributions from both the SM interactions and the additional sterile neutrino dipole portal.

In this work, we account for atomic binding effects by adopting the stepping approximation for the effective atomic charge $Z_{\mathrm{eff}}(T_e)$. Unlike the free electron approximation where all electrons are considered unbound, here an electron from a specific shell is treated as available for scattering only if the recoil energy exceeds its binding energy. This approach provides a more realistic description near the ionization thresholds \cite{Chen:2016eab}, and is typically applicable in the energy region considered in this analysis \cite{Kouzakov:2017hbc, Hsieh:2019hug}. Under this assumption, the effective atomic number is expressed as
\begin{align}
Z_{\mathrm{eff}}(T_e)=\sum_{\alpha}n_\alpha\theta(T_e - B_\alpha)
\end{align}
where $\theta(x)$ denotes the step function, $n_\alpha$ the number of electrons in atomic shell $\alpha$, and $B_\alpha$ the corresponding binding energy. The binding energies and electron occupation numbers for the xenon target material are taken from Ref.~\cite{xraydata:2009}.

The required minimum neutrino energy to produce a recoil energy with an outgoing sterile neutrino is
\begin{align}
E_{\nu}^{\text{min}} = \frac{T_e}{2}\left(1+\sqrt{1 + \frac{2m_e}{T_e}} \right) \left(1 + \frac{m_4^2}{2m_e T_e} \right),
\end{align}
while the maximum energy $E_{\nu,i}^{\text{max}}$ is taken from the neutrino flux endpoint. 

\begin{figure}[tbh]
	\centering	\includegraphics[scale=0.44]{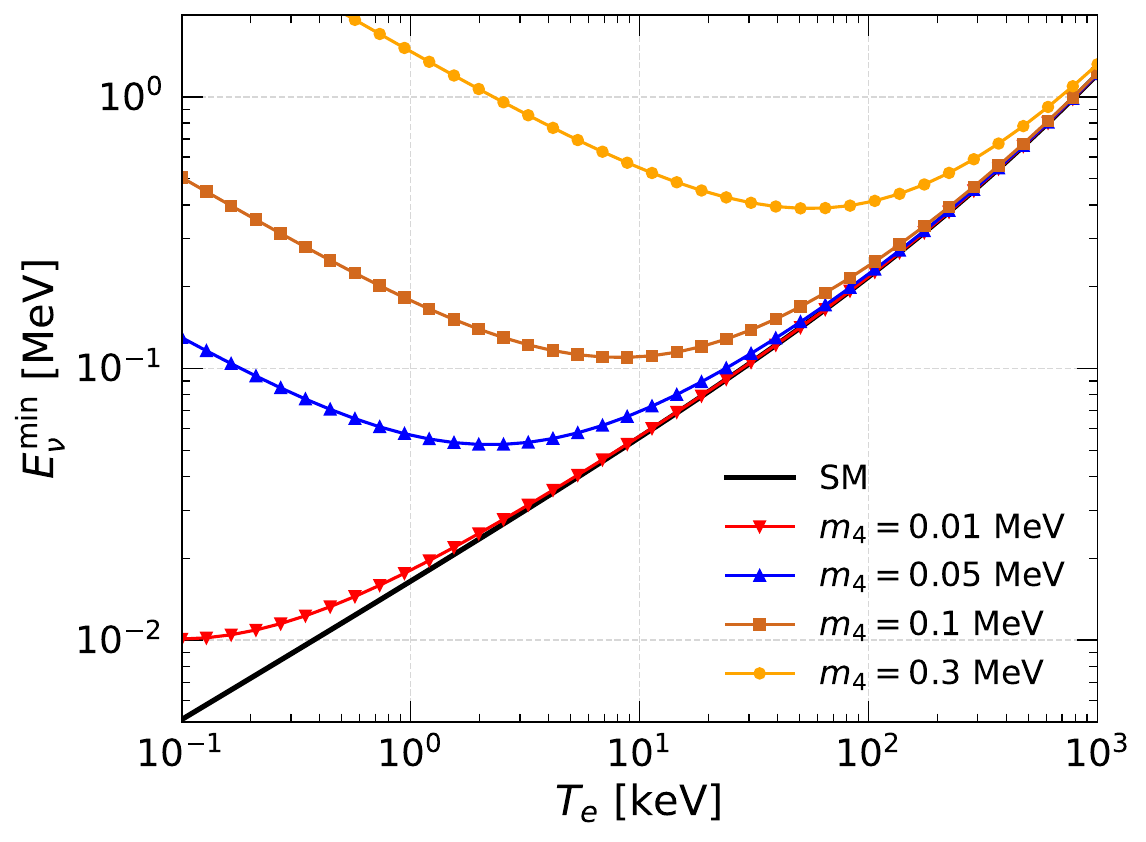}
	\caption{Behavior of the minimum neutrino energy with the electron recoil energy for different values of sterile neutrino masses. }
	\label{fig:Emin}
\end{figure}

The dependence of $E^\text{min}_\nu$ on $T_e$ is illustrated in Fig.~\ref{fig:Emin}. Several masses of the sterile neutrino are chosen to demonstrate the kinematic behavior. As expected, the minimum energy required to produce a recoil is linearly dependent on the increasing electronic recoil. At sufficiently high electron recoil energies, the kinematic threshold approaches the limit corresponding to the massless, active-only, neutrino case (solid-black line).
In the recoil energy region considered in this work, where $T_e \lesssim 30$ keV, different masses of the sterile neutrino noticeably modify the minimum required neutrino energy, thereby affecting the expected event rates. This dependency becomes particularly relevant in shaping the observable recoil spectrum, as heavier sterile neutrinos progressively suppress low-energy events due to the increasingly restrictive kinematic threshold. 

During their propagation from the Sun to the Earth, neutrinos experience flavor oscillations. Consequently, solar neutrinos arrive as a mixture of all possible flavors at the detector. 
The differential cross section needs to be multiplied by the corresponding oscillation probabilities
\begin{align}
	\left[\frac{d\sigma}{dT_{e}}\right]^{\nu e}_X = P_{ee} \left[\frac{d\sigma_{\nu_e}}{dT_{e}}\right]_X + \sum_{f=\mu,\tau} P_{ef} \left[\frac{d\sigma_{\nu_f}}{dT_{e}}\right]_X.
\end{align}
The $P_{ee}$ denotes the survival probability for $\nu_e$, and $P_{ef}$ for an initial $\nu_e$ converting into a $\nu_f$  flavor ($f=\mu, \tau$) at detection. These transition probabilities are given by $P_{e\mu}=(1-P_{ee})\cos^2\vartheta_{23}$ and $P_{e\tau}=(1-P_{ee})\sin^2\vartheta_{23}$, respectively. The electron neutrino survival probability, accounting for both vacuum and matter effects, is expressed as \cite{Maltoni:2015kca}
\begin{align}
	\begin{split}
		P_{ee} = & \cos^2 (\vartheta_{13}) {\cos^2 (\vartheta_{13}^m)}\left( \frac{1}{2} + \frac{1}{2} \cos (2\vartheta_{12}^m) \cos (2\vartheta_{12}) \right) \\ &+ \sin^2 (\vartheta_{13}) {\sin^2(\vartheta_{13}^m)},
		\label{Pee}
	\end{split}
\end{align}
where the neutrino mixing angles are given by $\vartheta_{12}, \vartheta_{13}$ and $ \vartheta_{23}$, while the superscript $m$ indicates quantities evaluated in the solar matter environment.
In our calculation, we account for the day–night asymmetry, coming from the regeneration of $\nu_e$ through coherent forward scattering with Earth’s matter during nighttime propagation. 
We use the normal mass ordering scenario for the neutrino oscillation parameters, which are taken from the global 3-flavor fit provided by NuFit-5.3 \cite{Esteban:2020cvm}, excluding the Super-Kamiokande atmospheric neutrino data. This particular choice is due to the lack of information available from the experiment and to avoid overfitting in the obtained results.

We calculate the predicted number of events in the $k$-th electron recoil energy bin using
\begin{align}
\begin{split}
R^k_X=&\varepsilon N_{t}\int_{T_e^k}^{T_e^{k+1}}dT_e \text{ } \mathcal{A}(T_e) \int_{0}^{T_e^{'\text{max}}}dT_e' \text{ } \mathcal{R}(T_e,T_e') \\ &\times\sum_{i=pp,^7\text{Be}} \left[\frac{dR}{dT'_{e}}\right]^i_X,
\end{split}
\end{align}
where $N_{t}$ is the number of target electrons per unit detector mass. The variables $T_e$ and $T_e'$ denote the reconstructed and true electron recoil energies, respectively. 
The functions $\mathcal{A}(T_e)$ account for the detector acceptance (efficiency), while $\mathcal{R}(T_e,T_e')$ is the energy resolution (smearing) function. 
The detector efficiency for PandaX-4T is taken from Ref.~\cite{PandaX:2024cic} and XENONnT from Ref.~\cite{XENON:2022ltv}. 
We consider a normalized Gaussian smearing function for the energy resolution with energy-dependent widths given by $\sigma =  0.073+0.173 T_e - 0.0065 T_e^2  + 0.00011 T_e^3$ (in keV) for PandaX-4T \cite{PandaX:2022ood} and $\sigma=0.31\sqrt{T_e} + 0.0037 T_e$ \cite{XENON:2020rca} for XENONnT. 
In order to compute the total event yield reported by each experiment, we need to multiply the differential event rate by the exposure factor $\varepsilon$. The PandaX-4T Run0 and Run1 datasets correspond to exposures of $198.9\text{ ton} \cdot \text{day}$ and $363.3\text{ ton} \cdot \text{day}$, respectively, while the XENONnT SR0 dataset has an exposure of $1.16 \text{ ton}\cdot\text{ year}$. The maximum allowed recoil energy is determined by kinematics and satisfies
\begin{align}
	T_e^{'\text{max}} = \frac{2E_\nu^2}{2E_\nu + m_e}.
\end{align}
This relation highlights that for a given neutrino energy, lighter targets yield higher maximum recoil energies, although in this context, the target electrons are identical for both detectors.

Both PandaX-4T \cite{PandaX:2024cic} and XENONnT \cite{XENON:2022ltv} have dual-phase (liquid + gas) xenon time projection chambers.  As a particle interacts in the liquid xenon, it produces two classes of measurable signals. The first signals is the prompted scintillation, labeled as S1, collected shortly after the particle interaction. The secondary delayed electroluminescence signals, labeled as S2, originated from ionization electrons freed by the recoil drift under an applied electric field to the liquid–gas interface. Both the S1 and S2 signals are the fundamental observables from which the measured energy, electron-equivalent or nuclear recoil energy, is constructed. Particularly, we focus on the recent low-energy electronic recoil datasets from these experiments.

\subsection{$\chi^2$ Function}
To derive constraints on the new physics parameter(s) of interest, denoted by $\mathcal{S}$, we employ a Poisson-based $\chi^2$ test statistic ~\cite{Baker:1983tu, Fogli:2002pt}, which is defined as
\begin{widetext}
\begin{align}
\begin{split}
\chi^2(\mathcal{S}) = \mathrm{min}_{(\alpha_i,\beta_j)} &2  \sum_{k=1}^{30} \biggl[  R_\text{exp}^{k}(\mathcal{S}; \alpha,\beta) - R_{obs}^{k}  + R_{obs}^{k} \ln \Bigg( \frac{R_{obs}^{k} }{R_\text{exp}^{k}(\mathcal{S}; \alpha,\beta) }\Bigg)  \biggr]  +\sum_{i}\left(\frac{\alpha_i}{\sigma_{\alpha_i}}\right)^2 +  \sum_{j}\left(\frac{\beta_j}{\sigma_{\beta_j}}\right)^2,
\end{split} \label{eq:chi2}
\end{align} 
\end{widetext}
where $R_{\text{obs}}^k$ and $R_{\text{exp}}^k$ represent the observed and expected event rates in the $k$-th energy bin, respectively. The expected rate consists of SM prediction $R^k_{\text{SM}}$, the contribution from the transition magnetic moment (TMM) denoted by $R^k_{\text{TMM}}$, and other background components $R_{\text{Bkg}}$. 
Explicitly, it is given by $R_{\text{exp}}^k(\mathcal{S};\alpha_i, \beta_j)=(1+\alpha_i)(R^k_{\text{SM}}+R^k_{\text{BSM}}(\mathcal{S}))$ +$  (1+\beta_j)R_{\text{Bkg},j}^k$ 
in which $\alpha_i$ and $\beta_j$ are nuisance parameters corresponding to the uncertainties on the solar neutrino fluxes and the individual background components, respectively. 
\begin{figure*}[!htb]
	\centering
	\includegraphics[scale=0.38]{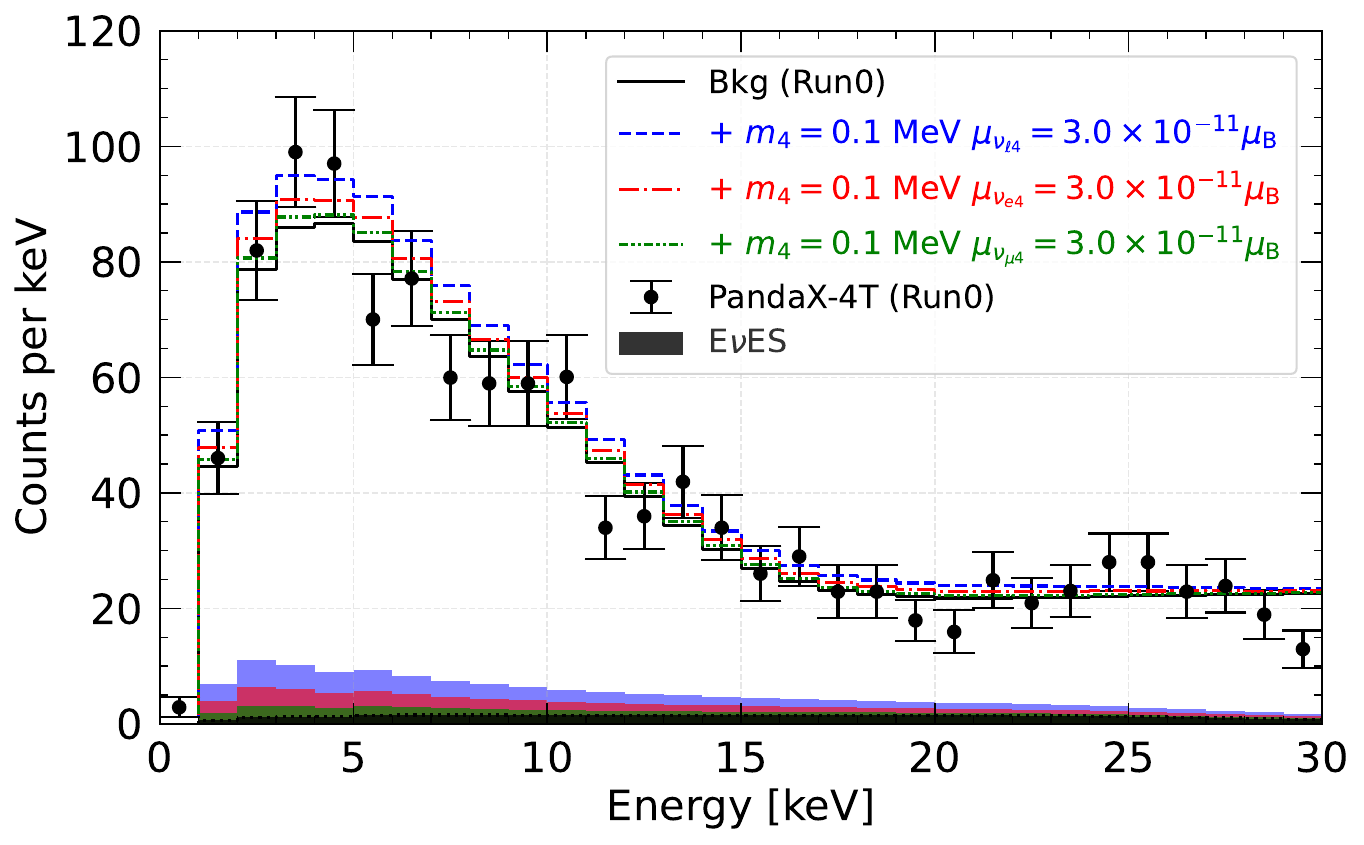}
	\includegraphics[scale=0.38]{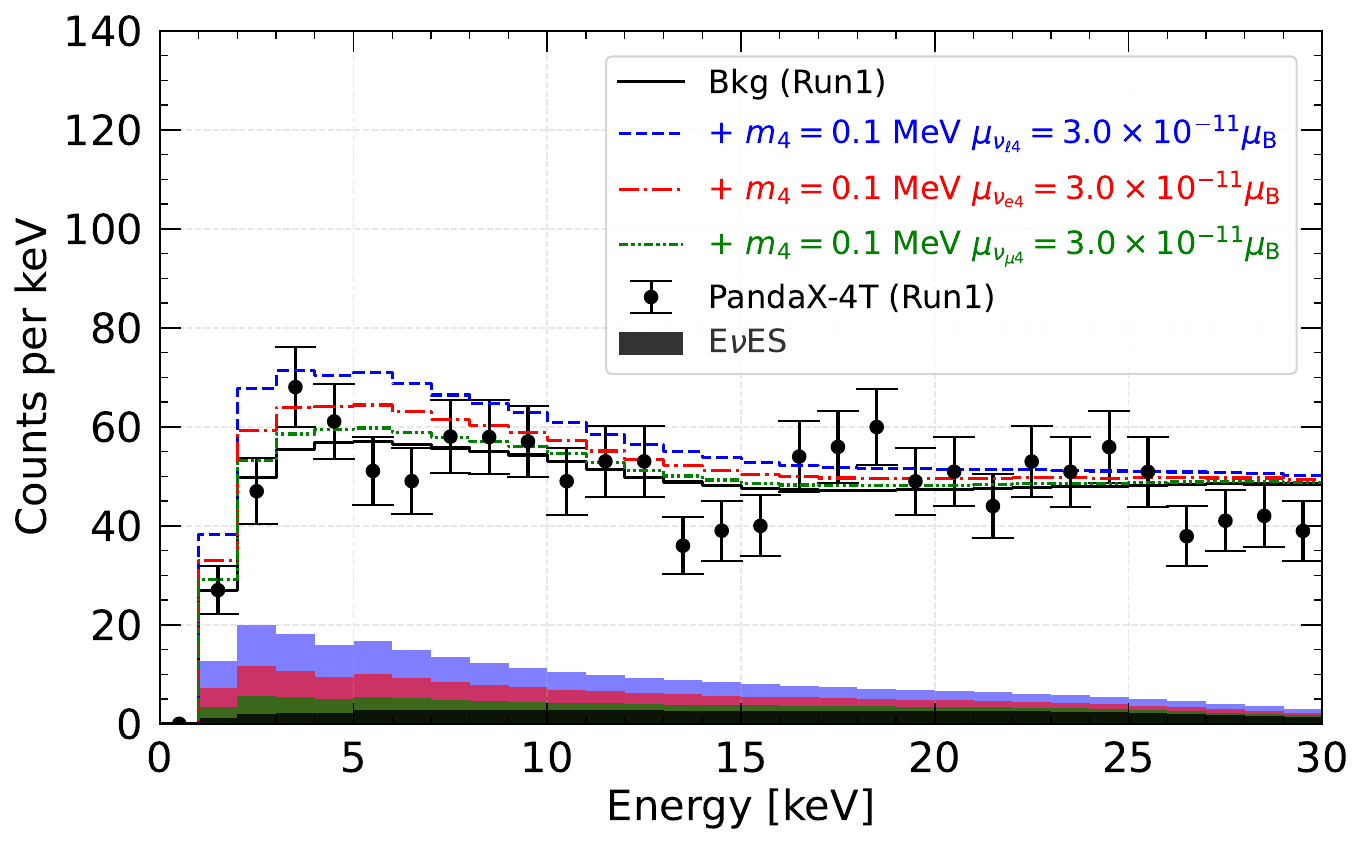}
	\\
	\hspace{8mm} (a) \hspace{71mm} (b)
	\\
	\includegraphics[scale=0.38]{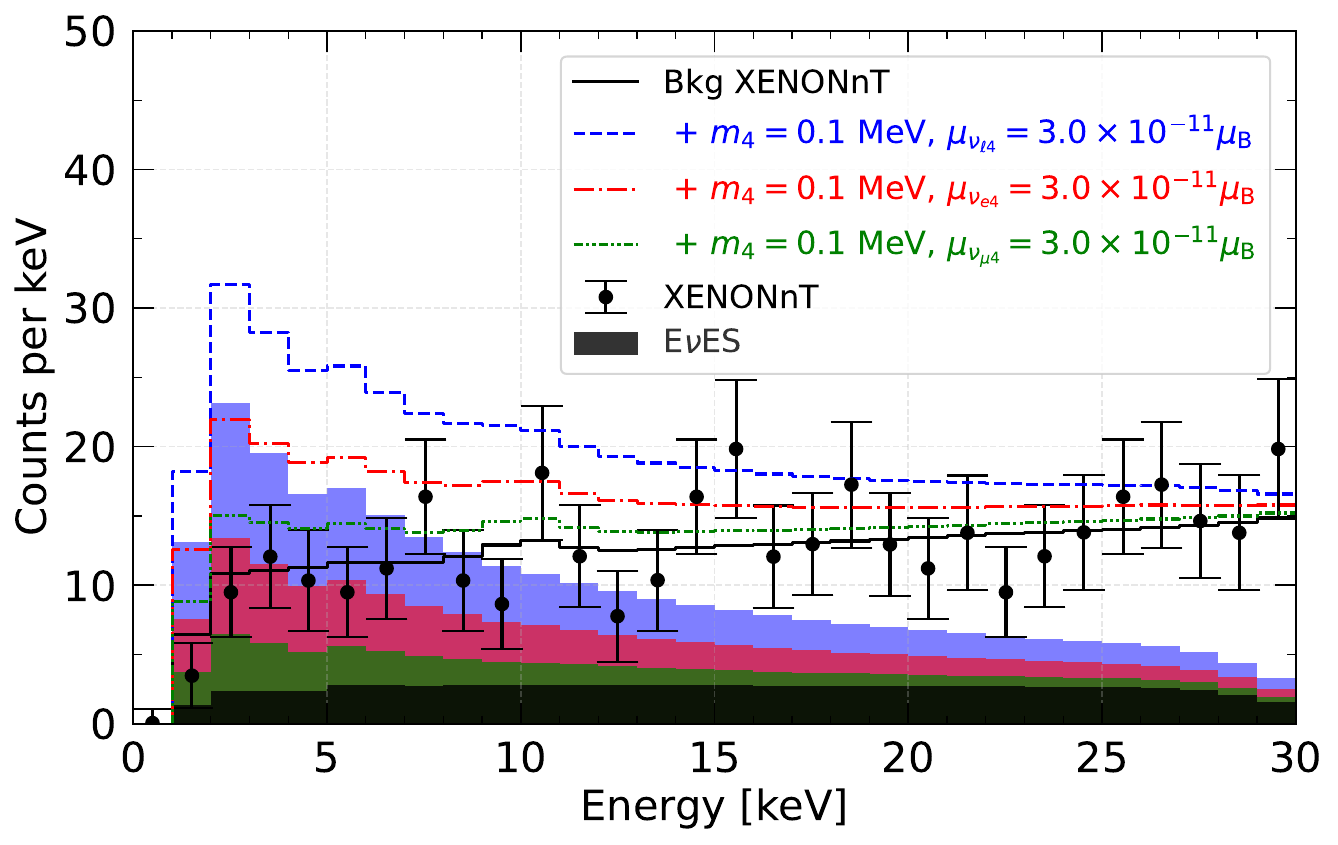}
	\\
	\hspace{6mm} (c)
	\\
	\vspace{-2mm}
	\caption{Predicted electron recoil spectra from E$\nu$ES in the presence of active–sterile transition magnetic moments, compared to the measured data from (a) PandaX-4T Run0, (b) PandaX-4T Run1, and (c) XENONnT. Individual contributions to the E$\nu$ES are shown as filled colored histograms, while their corresponding contributions to the total background are represented by empty, outlined histograms in the same colors.}
	\label{fig:rate_bsm}
\end{figure*}
The observed data points are taken from Fig.2 of Ref.\cite{PandaX:2024cic} for PandaX-4T, and from Fig.~4 (and Fig.5) of Ref.\cite{XENON:2022ltv} for XENONnT, considering recoil energies up to 30 keV. 
The uncertainty $\sigma_{\alpha}$ corresponds to the fractional uncertainties in the solar neutrino fluxes, as listed in Table 6 of Ref.~\cite{Vinyoles:2016djt}, while $\sigma_{\beta}$ represents the associated experimental background uncertainties.
The total predicted event rates—including both the SM and new physics contributions—are calculated using solar neutrino fluxes from Bahcall’s spectrum \cite{Bahcall:1989ks}, normalized according to the B16-GS98 Standard Solar Model \cite{Vinyoles:2016djt}.
\section{Results}\label{sec:results}

In this section, we present our numerical results for the active–sterile neutrino transition magnetic moment based on the analysis of data from direct detection experiments.  
We first examine the expected event spectra induced by neutrino-electron scattering. We then discuss our derived limits from analyzing recent electronic recoil signals from PandaX-4T (Run0 and Run1) and XENONnT. 

\subsection{Expected event spectra}
In Fig. \ref{fig:rate_bsm}, we show the electron recoil data from PandaX-4T and XENONnT together with the expected event rates in the presence of the active–sterile neutrino transition magnetic moment. The predicted E$\nu$ES signals induced by solar neutrinos are subtracted from the total background reported by each experiment. They are shown as solid black lines, exhibiting good agreement with the results published by the respective collaborations. 
In investigating the sensitivity to possible active–sterile neutrino transition magnetic moments, we have considered several benchmarks with sterile neutrino mass of $0.1 \text{ MeV}$ and transition magnetic moments of $3\times 10^{-11} \mu_B$.

\begin{figure*}[ht]
\centering
\includegraphics[scale=0.41]{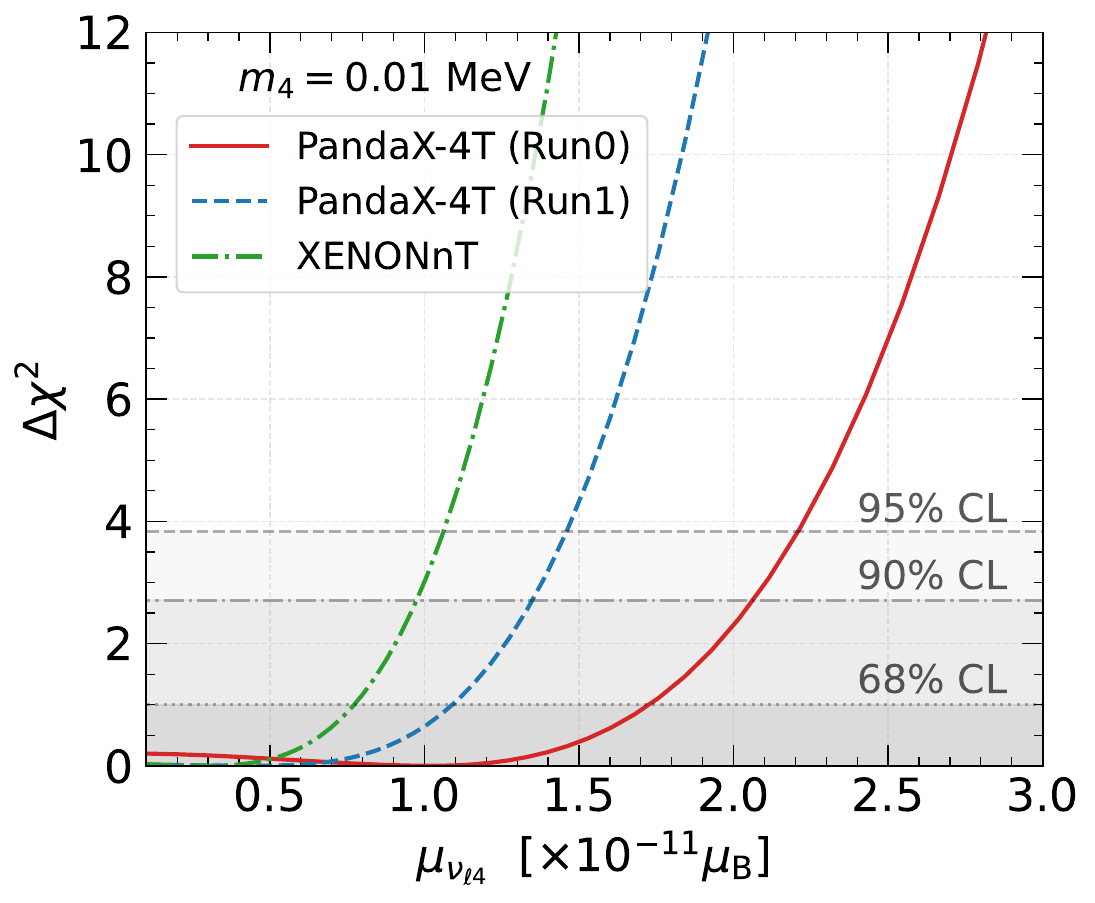}
\includegraphics[scale=0.41]{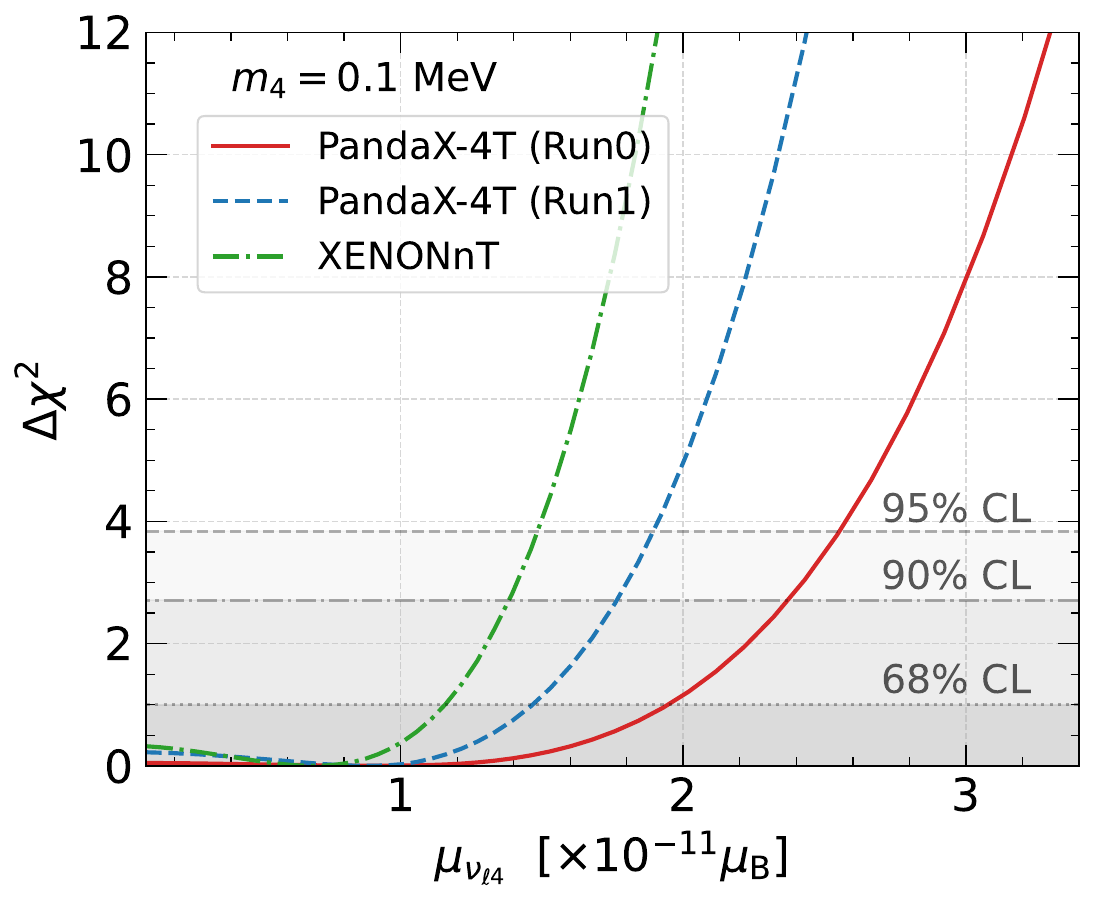}
	\\
\hspace{8mm} (a) \hspace{71mm} (b)
\\
\includegraphics[scale=0.41]{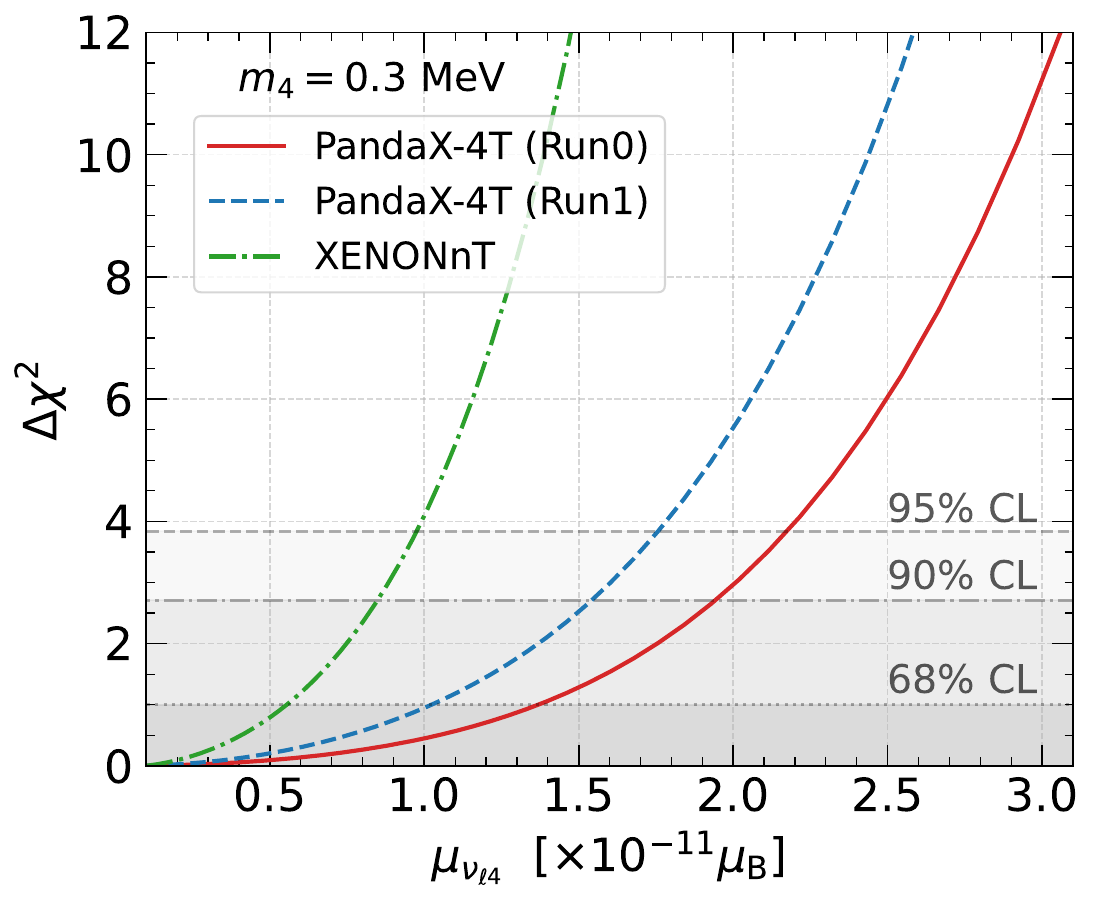}
	\\
\hspace{6mm} (c)
\\
\vspace{-2mm}
\caption{The likelihood profile of the effective active-transition magnetic moment $\mu_{\nu_{\ell 4}}$, obtained from PandaX-4T Run0, PandaX-4T Run1, and XENONnT data, for (a) $m_4=0.01 \text{ MeV}$, (b) $m_4=0.1 \text{ MeV}$, and (c) $m_4=0.3 \text{ MeV}$.}
\label{fig:1dof_eff}
\end{figure*}
\begin{figure*}[ht]
\centering
\includegraphics[scale=0.41]{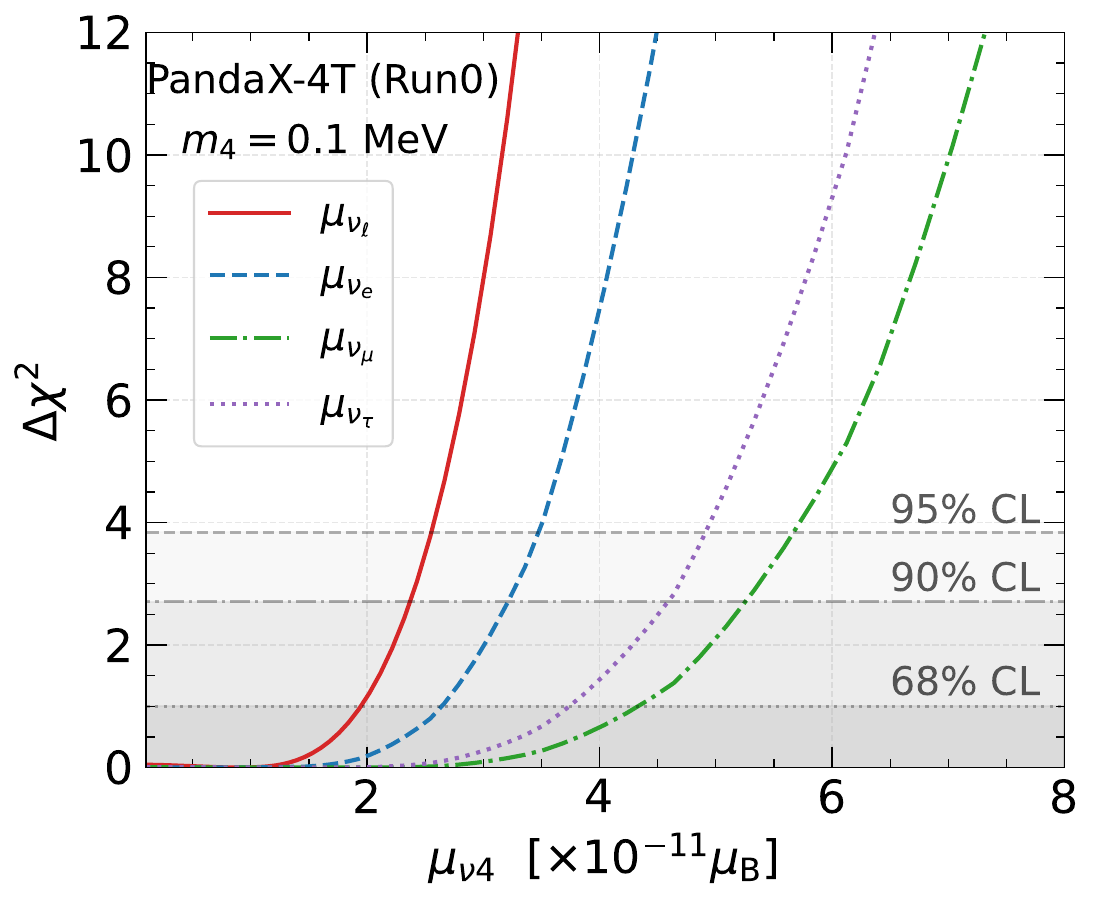}
\includegraphics[scale=0.41]{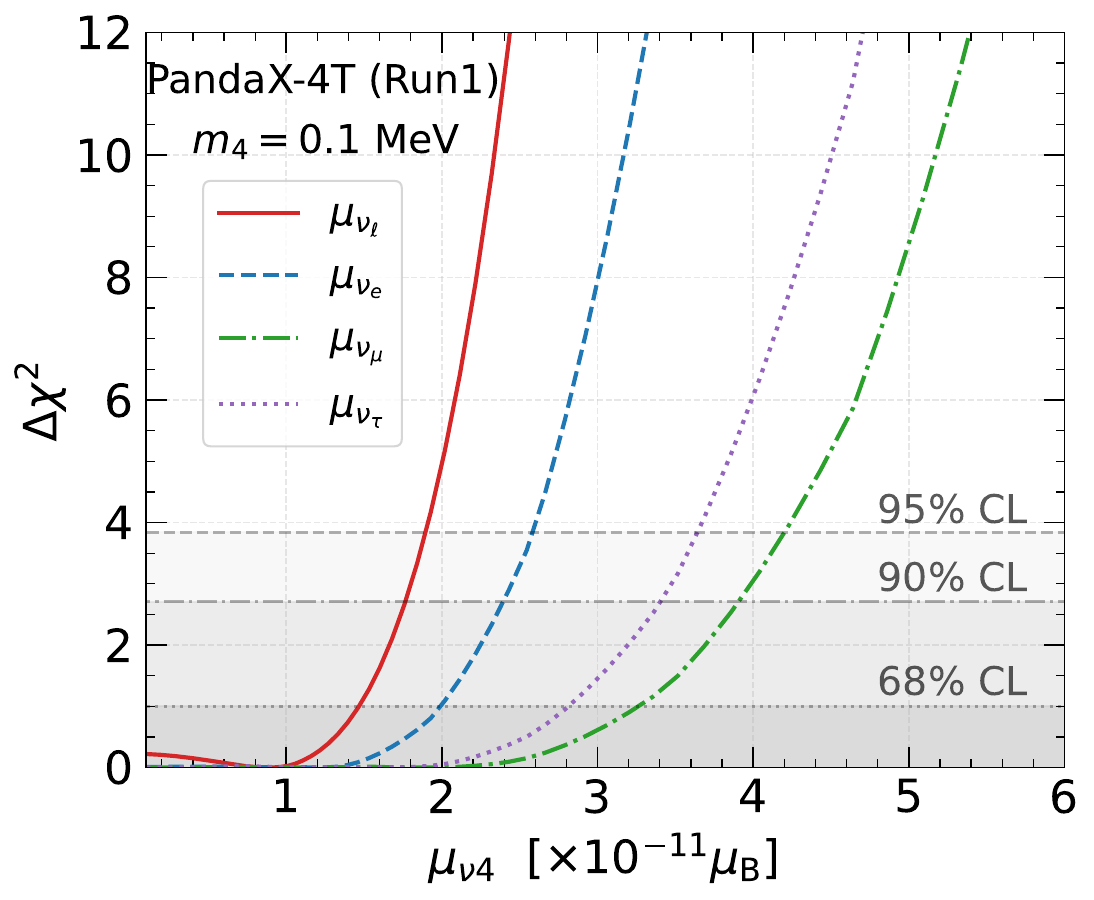}
	\\
\hspace{8mm} (a) \hspace{71mm} (b)
\\
\includegraphics[scale=0.41]{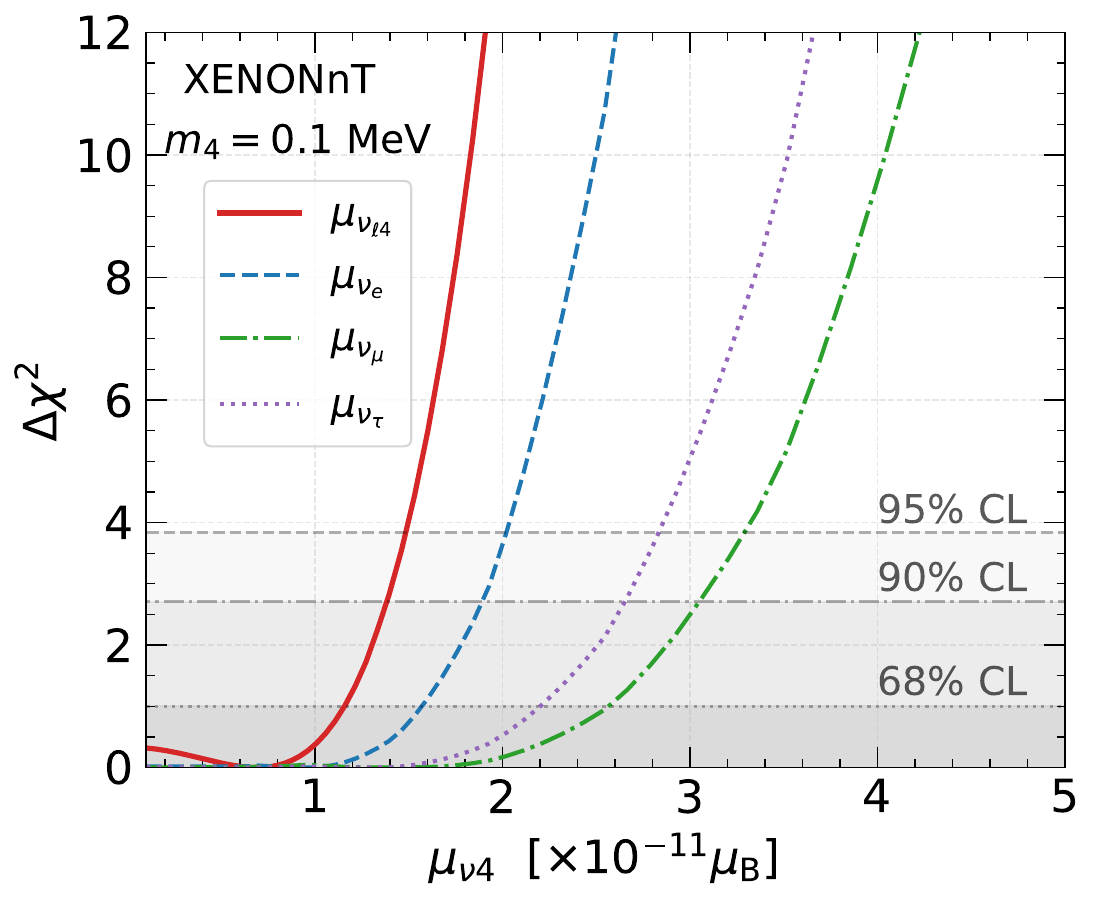}
	\\
\hspace{6mm} (c)
\\
\vspace{-2mm}
\caption{The likelihood profile of the active-transition magnetic moments obtained from (a) PandaX-4T Run0, (b) PandaX-4T Run1, and (c) XENONnT data for $m_4=0.1 \text{ MeV}$.}
\label{fig:1dof_01}
\end{figure*}
We consider both flavor-dependent and flavor-independent cases. 
In the flavor-dependent case, limits are individually derived for each active neutrino flavor, treating $\mu_{\nu_{e 4}}$, $\mu_{\nu_{\mu 4}}$, and $\mu_{\nu_{\tau 4}}$ as independent parameters, with marginalization performed over the remaining flavor components.
In the flavor-independent (universal coupling) case, $\mu_{\nu_{\ell 4}}=\mu_{\nu_{e 4}}=\mu_{\nu_{\mu 4}}=\mu_{\nu_{\tau 4}}$.
This scenario yields a single combined spectrum, incorporating contributions from all neutrino flavors weighted by their survival and transition probabilities. 
The predicted spectra for each flavor-specific transition magnetic moment are shown as colored lines in the figure to distinguish their contributions. 
As expected, the enhancement in event rates due to the presence of a transition magnetic moment is most prominent at low recoil energies.
This highlights the importance of improving experimental sensitivity in this region for future searches. The current datasets already provide useful constraints on probing scenarios involving neutrino electromagnetic interactions beyond the SM.
\subsection{Constraints on the sterile neutrino dipole portal}
We now discuss the limits we have derived on the active–sterile neutrino transition magnetic moments. As mentioned earlier, both flavor-dependent and flavor-independent scenarios are considered.
This classification is motivated by the properties of solar neutrinos, which contribute to the background components observed in direct detection experiments. In our analysis, we employ the most recent low-energy electron recoil datasets from the PandaX-4T (Run0 and Run1) and XENONnT experiments. These experiments offer unprecedented sensitivity to neutrino–electron scattering processes at recoil energies below 30 keV.

We also compare our derived limits with existing constraints reported in the literature. Previous studies have employed data from a variety of experimental platforms, including the COHERENT experiment at the Spallation Neutron Source (CsI and LAr detectors)~\cite{DeRomeri:2022twg}; reactor-based experiments utilizing both E$\nu$ES and CE$\nu$NS processes, such as CONUS+~\cite{DeRomeri:2025csu} and Dresden-II (FeF)~\cite{AristizabalSierra:2022axl}, as well as those relying solely on E$\nu$ES, such as TEXONO~\cite{Miranda2021}; the solar neutrino experiment BOREXINO~\cite{Brdar:2020quo}; and accelerator-based experiments including LSND (95\% CL)~\cite{Magill:2018jla} and CHARM-II~\cite{Coloma:2017ppo}. Additional constraints have also been obtained from other facilities such as MiniBooNE~\cite{Magill:2018jla}, NOMAD~\cite{Gninenko:1998nn} and IceCube~\cite{Coloma:2017ppo}. Furthermore, we include comparisons with indirect limits derived from sterile neutrino decay signatures using solar~\cite{Plestid:2020vqf} and atmospheric~\cite{Gustafson:2022rsz} neutrino data. For completeness, we also mention astrophysical bounds from SN1987A~\cite{Magill:2018jla} and cosmological observations~\cite{Brdar:2020quo}.

\begin{figure*}[ht]
	\centering
	\includegraphics[scale=0.44]{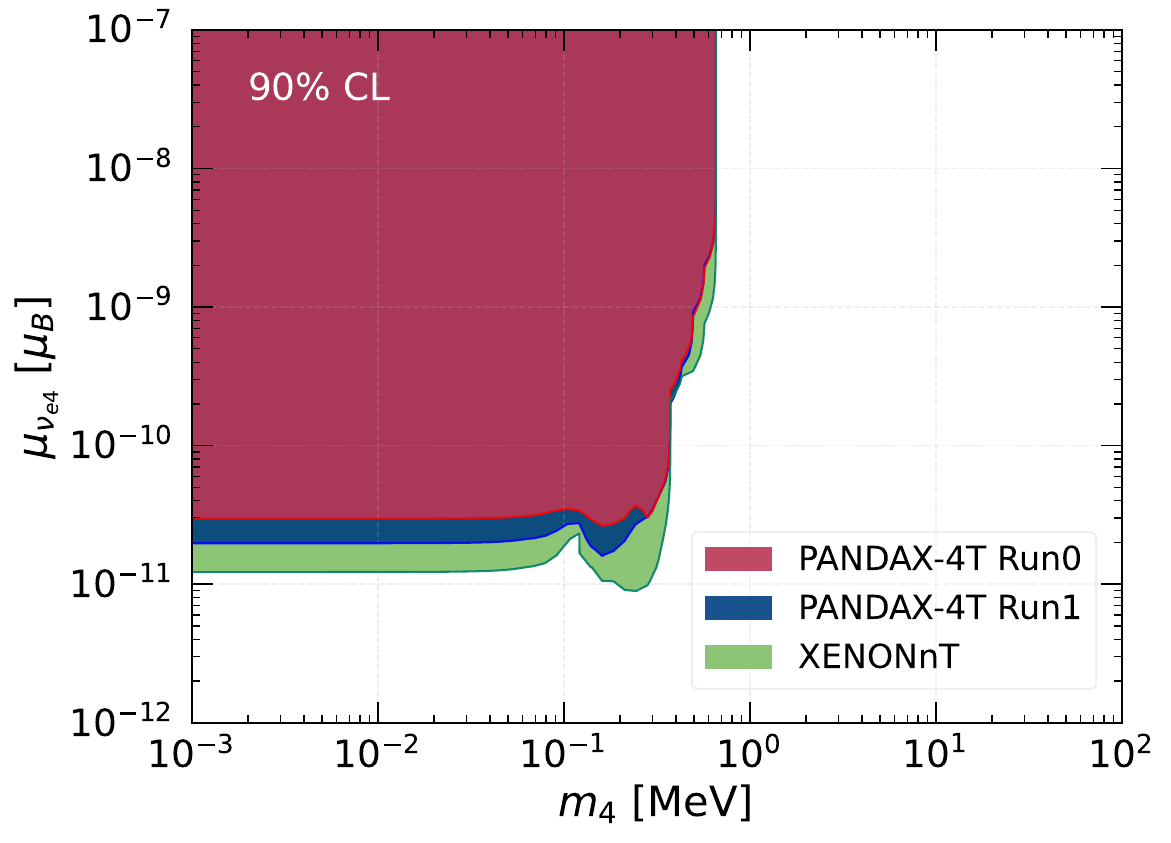}
    \includegraphics[scale=0.44]{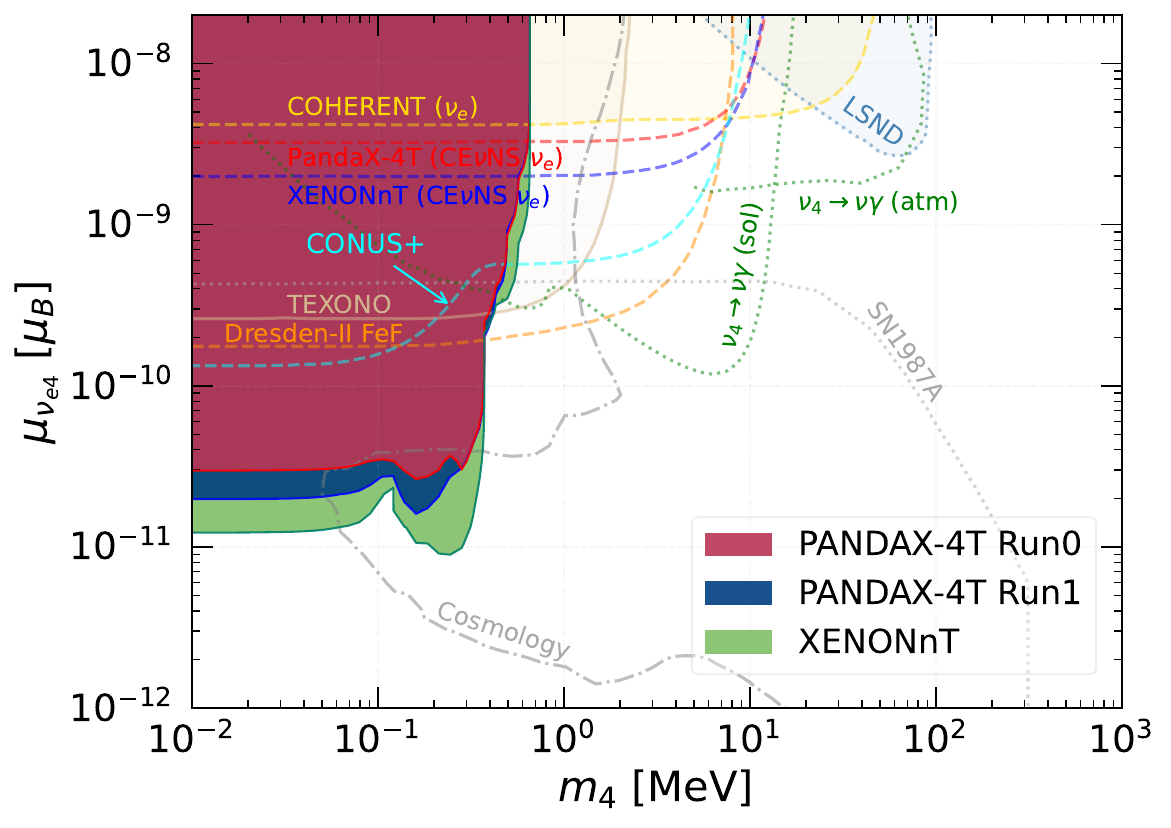}
    \\
 \hspace{10.4mm}	(a) \hspace{70mm} (b)
	\caption{(a) Exclusion regions with 90$\%$ CL (2 dof) on the plane of the $\mu_{\nu_{e4}}$ vs $m_4$ from PandaX-4T (Run0), (Run1) and XENONnT datasets, and (b) comparison with other available limits (see the text for details).}
	\label{fig:2dof_flav_e}
\end{figure*}
By setting a different mass scale for the effective sterile neutrino dipole portal, we present the 1 dof limits of the active-sterile neutrino magnetic moment in Fig.~\ref{fig:1dof_eff}. As can be seen from Fig.~\ref{fig:1dof_eff}(a), for $m_4=0.01$ MeV, we obtain the 90\% CL limits as $\mu_{\nu_{\ell 4}} \lesssim 2.06 \times 10^{-11} \mu_\text{B}$ for PandaX-4T Run0, $\mu_{\nu_{\ell 4}} \lesssim 1.35 \times 10^{-11} \mu_\text{B}$ for PandaX-4T Run1, and $\mu_{\nu_{\ell 4}} \lesssim 0.97 \times 10^{-11} \mu_\text{B}$ for XENONnT. In Fig.~\ref{fig:1dof_eff}(b), for $m_4=0.1$ MeV, we find the 90\% CL limits to be $\mu_{\nu_{\ell 4}} \lesssim 2.37 \times 10^{-11} \mu_\text{B}$ for PandaX-4T Run0, $1.76 \times 10^{-11} \mu_\text{B}$ for PandaX-4T Run1 and $\mu_{\nu_{\ell 4}} \lesssim 1.38 \times 10^{-11} \mu_\text{B}$ for XENONnT. Similarly, in Fig.~\ref{fig:1dof_eff}(c), for $m_4=0.3$ MeV, we determine the 90\% CL limits as $\mu_{\nu_{\ell 4}} \lesssim 1.94 \times 10^{-11} \mu_\text{B}$ for PandaX-4T Run0, $\mu_{\nu_{\ell 4}} \lesssim 1.54 \times 10^{-11} \mu_\text{B}$ for PandaX-4T Run1, and $\mu_{\nu_{\ell 4}} \lesssim 0.85 \times 10^{-11} \mu_\text{B}$ for XENONnT. These results clearly indicate the improvement in sensitivity provided by PandaX-4T Run1 data over Run0, 
while XENONnT maintains competitive limits across the considered mass values. 

Furthermore, we present the 1 dof limits for flavor-dependent cases in Fig.~\ref{fig:1dof_01}, where the mass of the sterile neutrino is fixed at $m_4=0.1$ MeV. From PandaX-4T Run0, we obtain the 90\% CL limits as $\mu_{\nu_{\ell 4}} \lesssim 2.37 \times 10^{-11} \mu_\text{B}$, $\mu_{\nu_{e 4}} \lesssim 3.21 \times 10^{-11} \mu_\text{B}$, $\mu_{\nu_{\mu 4}} \lesssim 5.25 \times 10^{-11} \mu_\text{B}$, and $\mu_{\nu_{\tau 4}} \lesssim 4.59 \times 10^{-11} \mu_\text{B}$. For PandaX-4T Run1, we find the limits to be $\mu_{\nu_{\ell 4}} \lesssim 1.76 \times 10^{-11} \mu_\text{B}$, $\mu_{\nu_{e 4}} \lesssim 2.39 \times 10^{-11} \mu_\text{B}$, $\mu_{\nu_{\mu 4}} \lesssim 3.91 \times 10^{-11} \mu_\text{B}$, and $\mu_{\nu_{\tau 4}} \lesssim 3.40 \times 10^{-11} \mu_\text{B}$. From XENONnT data, the corresponding limits are $\mu_{\nu_{\ell 4}} \lesssim 1.38 \times 10^{-11} \mu_\text{B}$, $\mu_{\nu_{e 4}} \lesssim 1.89 \times 10^{-11} \mu_\text{B}$, $\mu_{\nu_{\mu 4}} \lesssim 3.05 \times 10^{-11} \mu_\text{B}$, and $\mu_{\nu_{\tau 4}} \lesssim 2.65 \times 10^{-11} \mu_\text{B}$. A summary of these results is provided in Table~\ref{tab:1dof_flavdep} for clarity and comparison.
We clearly observe that, for flavor-dependent case, the sensitivity to $\mu_{\nu_{e 4}}$ is consistently stronger than those for $\mu_{\nu_{\mu 4}}$ and $\mu_{\nu_{\tau 4}}$ across all datasets, which is a direct consequence of the dominant contribution of electron-flavor solar neutrinos to electrons in elastic scattering processes. 
Among the three datasets, PandaX-4T Run1 provides notable improvements over Run0 for all flavor cases, while XENONnT continues to set the most stringent constraints, particularly for $\mu_{\nu_{e 4}}$ and $\mu_{\nu_{\ell 4}}$ couplings.
\begin{table}[ht]
\caption{Summary of the derived 90\% CL limits (1 dof) on the transition magnetic moments for $m_4=0.1$ MeV, obtained from the PandaX-4T and XENONnT datasets.
}
\begin{center}
\begin{ruledtabular}
\begin{tabular}{ c c c c c c c}
\multirow{2}{1.5cm}{TMM $\times 10^{-11}\mu_\mathrm{B}$} & \multicolumn{2}{c}{PandaX-4T} & \multicolumn{1}{c}{XENONnT}
			\\
			\cline{2-3}
 &  Run0 & Run1 & \\
 \hline
$\mu_{\nu_{\ell 4}} $ & $\lesssim 2.37 $ & $\lesssim 1.76$ & $\lesssim 1.38 $ \\ 
$\mu_{\nu_{e 4}} $ & $\lesssim 3.21 $ & $\lesssim 2.39 $ & $\lesssim 1.89 $ \\ 
$\mu_{\nu_{\mu 4}} $ & $\lesssim 5.25 $ & $\lesssim3.91 $ & $\lesssim 3.05$ \\ 
$\mu_{\nu_{\tau 4}} $ & $\lesssim 4.59 $ & $\lesssim3.40$ & $\lesssim 2.65$ \\ 
\end{tabular}
\end{ruledtabular}
\end{center}
\label{tab:1dof_flavdep}
\end{table}

We now discuss the 2 dof results at 90\% CL for the electron-flavor case as shown in Fig.~\ref{fig:2dof_flav_e}(a). 
For $m_4\lesssim 0.03$ MeV, the derived upper limits are $\mu_{\nu_{e 4}} \lesssim 2.97 \times 10^{-11} \mu_\text{B}$ for PandaX-4T Run0, $\mu_{\nu_{e 4}} \lesssim 1.97 \times 10^{-11} \mu_\text{B}$ for PandaX-4T Run1, and $\mu_{\nu_{e 4}} \lesssim 1.22 \times 10^{-11} \mu_\text{B}$ for XENONnT in the low-mass region. We observe a substantial improvement from PandaX-4T Run0 to Run1, where the upper limit is reduced by more than a factor of two. This enhancement is primarily attributed to the increased exposure and improved background suppression in Run1. Meanwhile, XENONnT delivers the most stringent constraint in this channel, slightly outperforming PandaX-4T Run1, which reflects the advantage of its larger exposure and lower background levels in the relevant low-energy region. 
In Figure~\ref{fig:2dof_flav_e}(b), we compare our results with previous experimental limits, highlighting the significant improvements achieved by recent datasets—particularly those from PandaX-4T Run 1 and XENONnT. Notably, our derived limits
are more stringent than prior constraints in the sub-MeV mass range.
Our results are approximately two orders of magnitude more stringent than the limits obtained from CE$\nu$NS processes, such as those reported by COHERENT and from solar neutrino–induced backgrounds in PandaX-4T and XENONnT.
In comparison to constraints from nuclear reactor experiments—CONUS+, TEXONO, and Dresden-II FeF—the discrepancy is reduced to about one order of magnitude. However, at mass scales $m_4 \gtrsim 1~\mathrm{MeV}$, our results do not surpass the existing limits set by LSND and $\nu_4 \rightarrow \nu \gamma$ decay searches. Additionally, our constraints partially overlap with the parameter space excluded by SN1987A and cosmological observations. 
\begin{figure*}[ht]
	\centering
	\includegraphics[scale=0.44]{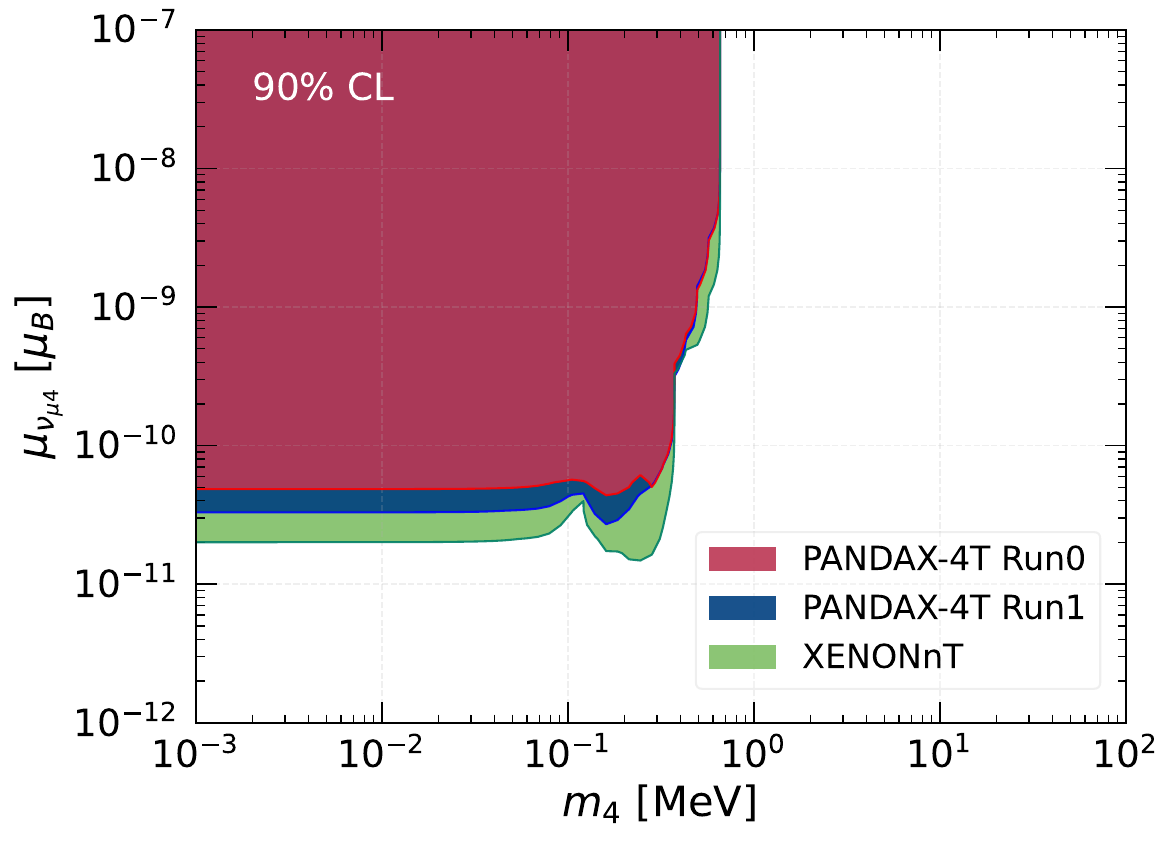}
    \includegraphics[scale=0.44]{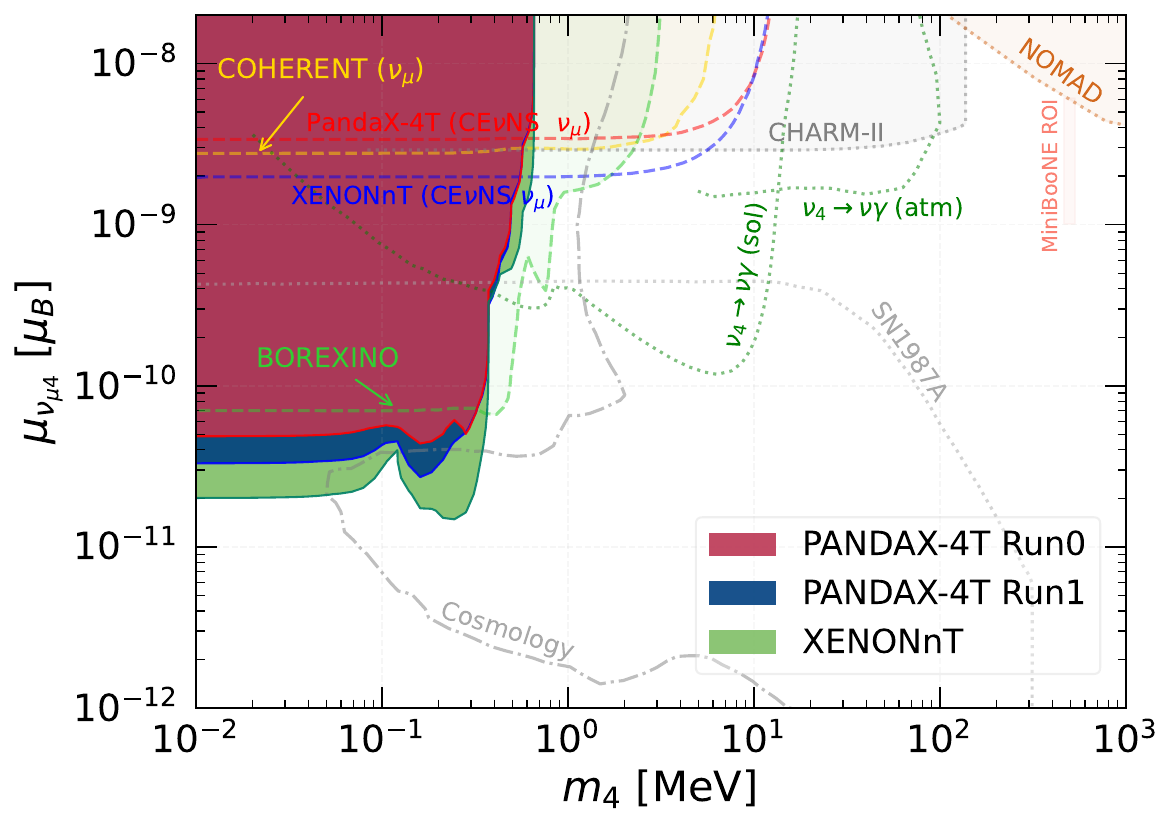}
    \\
 \hspace{10.4mm}	(a) \hspace{70mm} (b)
	\caption{(a) Exclusion regions with 90$\%$ CL (2 dof) on the plane of the $\mu_{\nu_{\mu 4}}$ vs $m_4$ from PandaX-4T (Run0), (Run1) and XENONnT datasets, and (b) comparison with other available limits (see the text for details).}
	\label{fig:2dof_flav_mu}
\end{figure*}

\begin{figure*}[ht]
	\centering
	\includegraphics[scale=0.44]{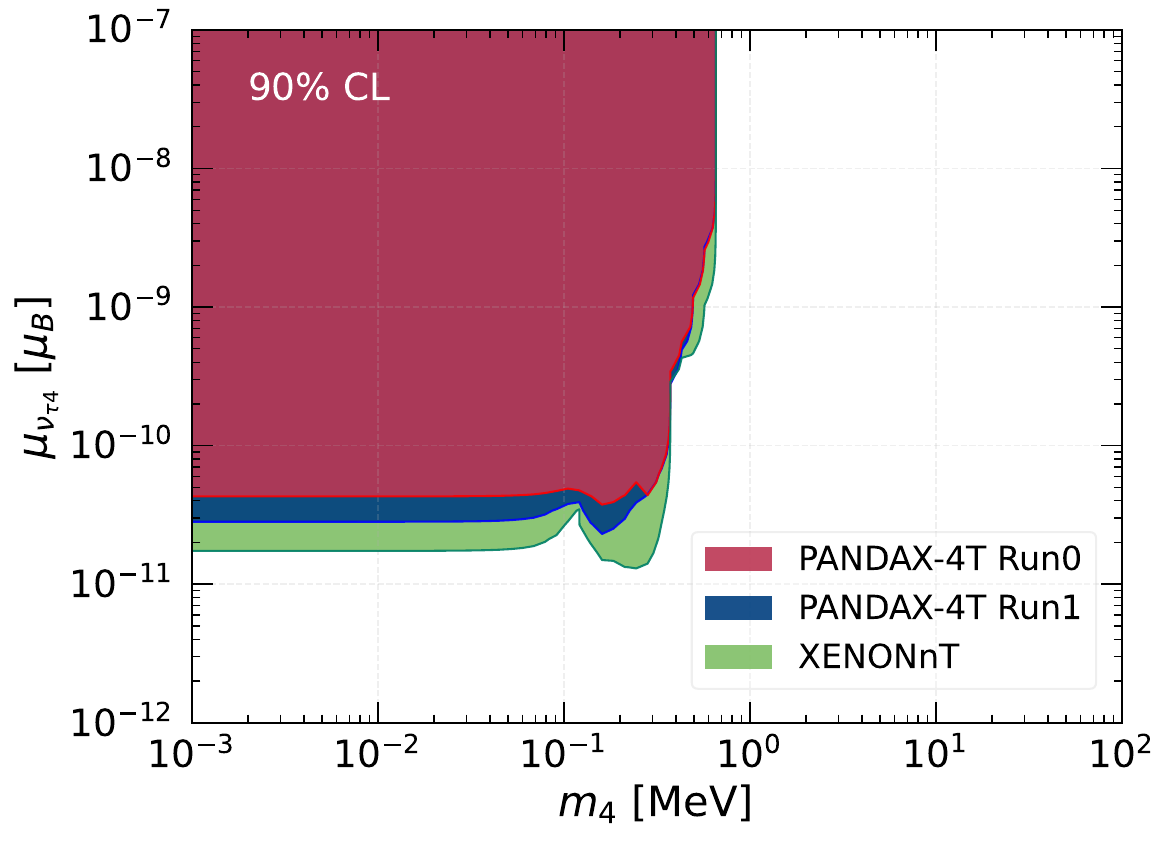}
    \includegraphics[scale=0.44]{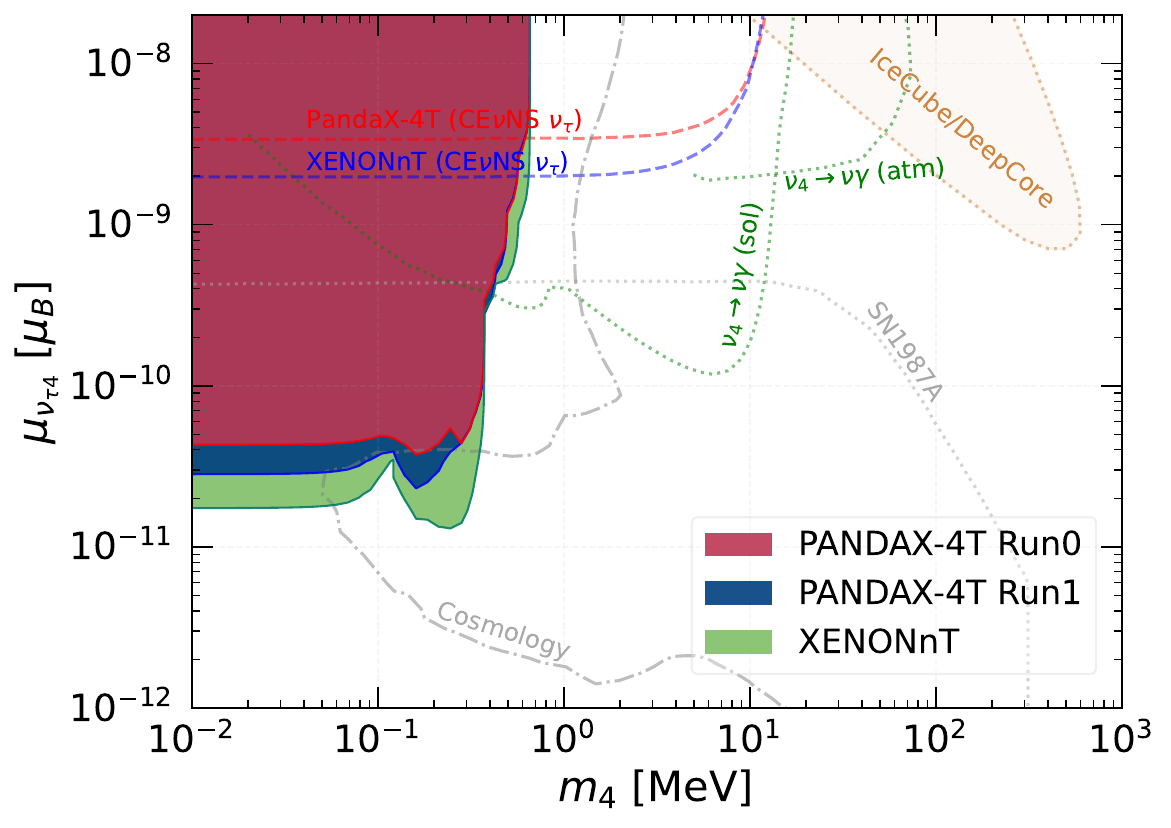}
    \\
 \hspace{10.4mm}	(a) \hspace{70mm} (b)
	\caption{(a) Exclusion regions with 90$\%$ CL (2 dof) on the plane of the $\mu_{\nu_{\tau 4}}$ vs $m_4$ from PandaX-4T (Run0), (Run1) and XENONnT datasets, and (b) comparison with other available limits (see the text for details).}
	\label{fig:2dof_flav_tau}
\end{figure*}
%
\begin{figure*}[ht]
	\centering
		\includegraphics[scale=0.44]{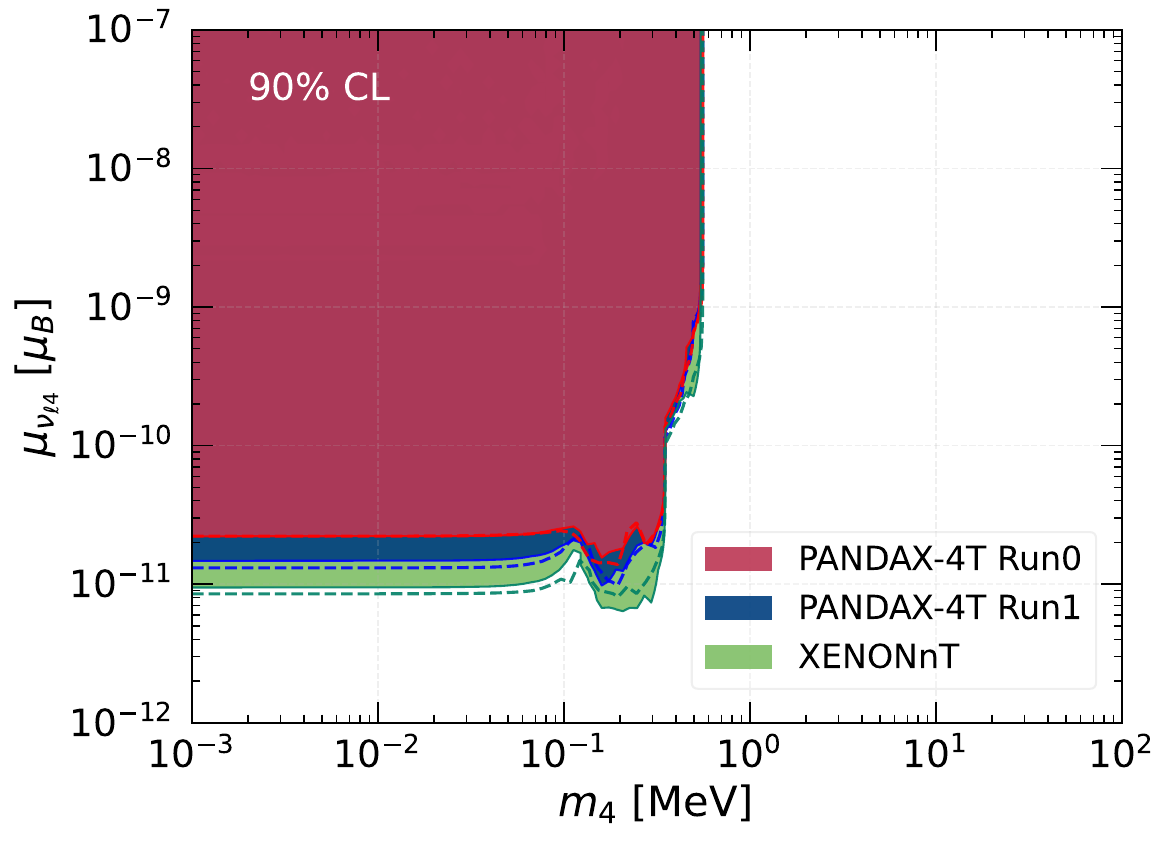}
    \includegraphics[scale=0.44]{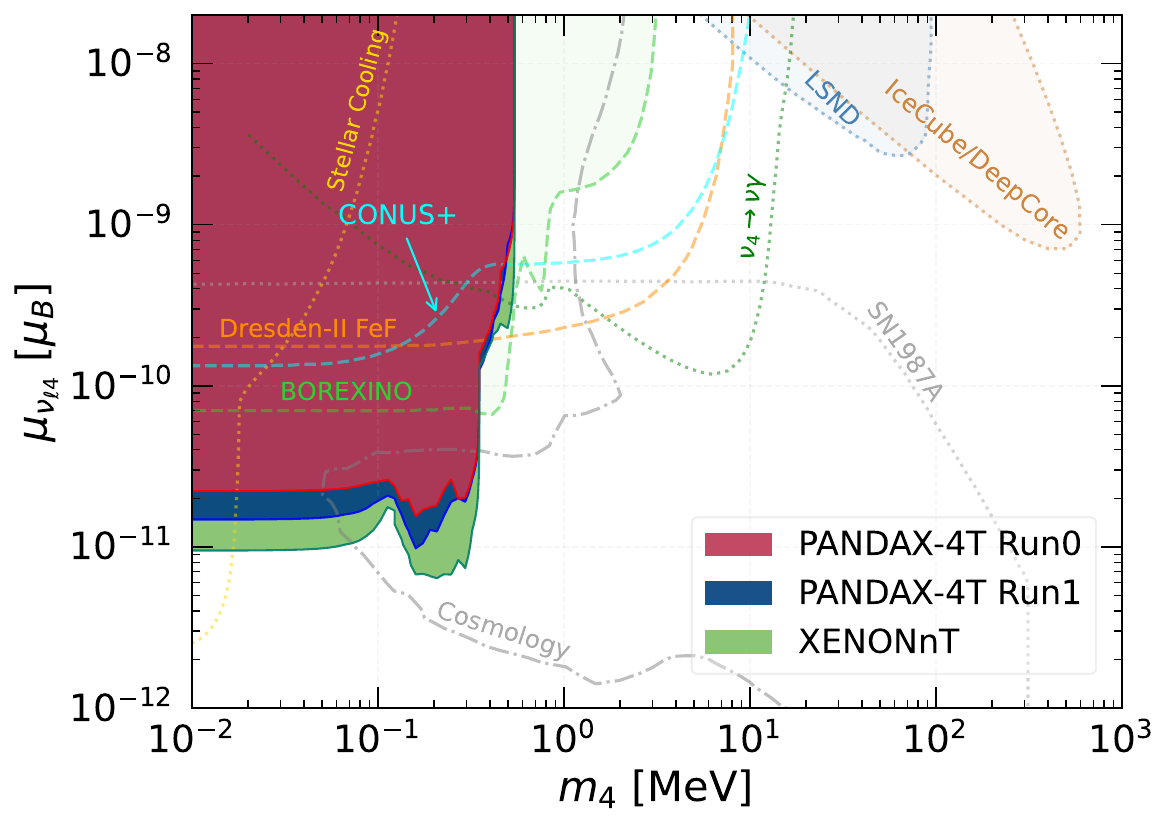}
    \\
 \hspace{10.4mm}	(a) \hspace{70mm} (b)
	\caption{(a) Exclusion regions with 90$\%$ CL (2 dof) on the plane of the $\mu_{\nu_{\ell 4}}$ vs $m_4$ from PandaX-4T (Run0), (Run1) and XENONnT datasets, together with results from more conservative analysis, and (b) comparison with other available limits (see the text for details).}
	\label{fig:2dof_eff}
\end{figure*}

For the muon-flavor case, we present the 2 dof results at 90\% CL in Fig.\ref{fig:2dof_flav_mu}(a). The derived upper limits in the region of $m_4 \lesssim 0.03$ MeV are $\mu_{\nu_{\mu 4}} \lesssim 4.86 \times 10^{-11} \mu_\text{B}$ for PandaX-4T Run0, $\mu_{\nu_{\mu 4}} \lesssim 3.30 \times 10^{-11} \mu_\text{B}$ for PandaX-4T Run1, and $\mu_{\nu_{\mu 4}} \lesssim 2.01 \times 10^{-11} \mu_\text{B}$ for XENONnT. We observe a clear improvement in sensitivity from Run0 to Run1, with the upper limit reduced by more than a factor of two. Additionally, XENONnT provides a slightly stronger bound than PandaX-4T Run1, confirming its competitive performance in this channel. A comparison with previous limits is shown in Fig.\ref{fig:2dof_flav_mu}(b). Apart from the already mentioned limits in the electron-flavor case, here we also superimpose additional limits from the solar neutrino experiment and collider facilities. We can see the PandaX-4T Run0 limit is about as sensitive as the BOREXINO limit, while the PandaX-4T Run1 and XENONnT limits provide a few times more stringent results than from this solar neutrino experiment. Meanwhile, limits from CHARM-II, MiniBooNE, and NOMAD dominate at the high-mass region. Regarding the already mentioned limits, we can see similar behavior to the electron-flavor case.

For the tau-flavor case, the 2 dof results at 90\% CL are shown in Fig.\ref{fig:2dof_flav_tau}(a). The obtained upper limits are $\mu_{\nu_{\tau 4}} \lesssim 4.30 \times 10^{-11} \mu_\text{B}$ for PandaX-4T Run0, $\mu_{\nu_{\tau 4}} \lesssim 2.82 \times 10^{-11} \mu_\text{B}$ for PandaX-4T Run1, and $\mu_{\nu_{\tau 4}} \lesssim 1.74 \times 10^{-11} \mu_\text{B}$ for XENONnT, all in the region of $m_4 \lesssim 0.03$ MeV. Analogous to electron-flavor and muon-flavor cases, we observe a substantial improvement between Run0 and Run1, while XENONnT delivers the most stringent constraint. This result emphasizes the increasing sensitivity of recent direct detection experiments to the tau-flavor sector, which traditionally suffers from weaker constraints due to the absence of direct tau neutrino sources at these energies. The comparison with existing bounds is provided in Fig.\ref{fig:2dof_flav_tau}(b). In the currently considered parameter space, there are but a few available limits. We can see the induced solar neutrino CE$\nu$NS processes of PandaX-4T and XENONnT are surpassed for $m_4 \lesssim 0.4$ MeV, while these limits dominate the high-mass region up to~10 MeV. Moreover, the sterile-neutrino decay and IceCube/DeepCore bounds occupy the much larger mass region up to the sub-GeV scale.

Lastly, we report the 2 dof results at 90\% CL for the flavor-independent case in Fig.\ref{fig:2dof_eff}(a). The corresponding upper limits for $m_4 \lesssim 0.03$ MeV are $\mu_{\nu_{\ell 4}} \lesssim 2.22 \times 10^{-11} \mu_\text{B}$ for PandaX-4T Run0, $\mu_{\nu_{\ell 4}} \lesssim 1.47 \times 10^{-11} \mu_\text{B}$ for PandaX-4T Run1, and $\mu_{\nu_{\ell 4}} \lesssim 0.95 \times 10^{-11} \mu_\text{B}$ for XENONnT \footnote{In the limit of low sterile neutrino mass, our derived upper bound is weaker than the experimental limit of $\mu_{\nu} < 6.4 \times 10^{-12} \mu_\text{B}$  reported by XENONnT \cite{XENON:2022ltv}, which is expected given that our analysis does not include full detector response simulations.}. Similar with the flavor-dependent cases, we observe significant improvements in the limits from Run0 to Run1, while XENONnT provides the tightest constraint overall. In addition, we perform a more conservative analysis by assigning a single nuisance parameter to the backgrounds, the results of which are indicated by the dashed lines. 
We observe that the conservative approach can yield more stringent constraints by up to a factor of two. 
This difference highlights the importance of our nominal methodology: modeling individual background components allows the analysis to leverage distinct spectral shape information, which is essential for maximizing experimental sensitivity.
A detailed comparison with earlier results of flavor-dependent cases is presented in Fig.\ref{fig:2dof_eff}(b). In general, similar behavior from previous results is found. Supplementary limit from stellar cooling \cite{Brdar:2020quo} process is added, indicating that our obtained limits dominate the sub-MeV mass scale and are about one order of magnitude away from covering this bound. Overall, the recent low-energy electronic recoil data from direct-detection facilities could provide limit enhancements on the transition magnetic moment compared to other facilities. This comparison highlights the progressive tightening of bounds achieved by the latest datasets.

\begin{table}[ht]
\caption{
Summary of the derived 90\% CL limits (2 dof) on the transition magnetic moments, obtained from the PandaX-4T and XENONnT datasets.
}
\begin{center}
\begin{ruledtabular}
\begin{tabular}{ c c c c c c c}
\multirow{2}{1.5cm}{TMM $\times 10^{-11}\mu_\mathrm{B}$ } & \multicolumn{2}{c}{PandaX-4T} & \multicolumn{1}{c}{XENONnT}
			\\
			\cline{2-3}
 &  Run0 & Run1 & 
	\\	\hline
$\mu_{\nu_{\ell 4}} $ & $\lesssim 2.22 $ & $\lesssim 1.47$ & $\lesssim 0.95$ \\ 
$\mu_{\nu_{e 4}} $ & $\lesssim 2.97 $ & $\lesssim 1.97$ & $\lesssim 1.22 $ \\ 
$\mu_{\nu_{\mu 4}}$ & $\lesssim 4.86 $ & $\lesssim 3.30 $ & $\lesssim 2.01 $ \\ 
$\mu_{\nu_{\tau 4}}$ & $\lesssim 4.30 $ & $\lesssim 2.82 $ & $\lesssim 1.74 $ \\ 
\end{tabular}
\end{ruledtabular}
\end{center}
\label{tab:2dof_flavdep}
\end{table}
We summarize our 2 dof 90\% CL limits in Table~\ref{tab:2dof_flavdep}, which illustrates the comparative sensitivity of the considered experiments. We note that a combined analysis using datasets from some DD experiments, including earlier PandaX data, was previously performed in Ref.~\cite{DeRomeri:2024hvc}.
Our results demonstrate an improvement over the limits reported in that work. 
Beyond current experimental constraints, future prospects for probing active–sterile TMMs have been actively explored in the literature. In particular, Ref.\cite{Brdar:2020quo} provides projected sensitivities for the next-generation multi-ton direct detection experiment DARWIN \cite{DARWIN:2016hyl}, demonstrating its potential to significantly improve upon existing limits. 
Similarly, in our previous work~\cite{Demirci:2024vzk}, we performed a sensitivity analysis using the projected specifications of the CDEX-50 experiment~\cite{CDEX:2023vvc} for CE$\nu$NS processes, highlighting its complementary reach in the low-energy neutrino sector.
Additionally, Ref.\cite{Ismail:2021dyp} investigates expected sensitivities at future forward neutrino detectors such as FLArE \cite{Batell:2021blf}, designed to probe TeV-scale neutrino interactions at the LHC. Their projections for sterile neutrino dipole portals at high neutrino energies further complement the low-energy direct detection constraints.
Overall, our results reinforce the complementarity of the latest direct detection data with other experimental approaches in constraining active–sterile neutrino transition magnetic moments. The growing precision and exposure of current-generation direct detection experiments, combined with future facilities, promise continued progress in probing neutrino electromagnetic properties beyond the SM.

\section{Conclusions}\label{sec:conc}
Motivated by the availability of recent low-energy electron recoil data from direct detection experiments, we have derived stringent constraints on the active–sterile neutrino transition magnetic moments. The improved sensitivity of facilities such as PandaX-4T and XENONnT to low-energy electron recoils has enabled the observation of solar neutrino fluxes with unprecedented precision, thereby providing an ideal platform to probe the sterile neutrino dipole portal.
We have explored both flavor-dependent and flavor-independent scenarios of active–sterile transition magnetic moments, incorporating the effects of neutrino oscillations and flavor conversion in the Sun and Earth. 

In this work, we have considered the contributions of both $pp$ and $^7$Be solar neutrino flux components in driving the up-scattering process via a transition magnetic moment into a sterile neutrino state. The analysis includes the day–night asymmetry effect arising from the matter-induced oscillation of neutrinos traversing the Earth, ensuring a comprehensive treatment of solar neutrino propagation effects.
We have presented limits at both 1 dof and 2 dof of the sterile neutrino dipole portal, providing a robust statistical interpretation of the results. Our derived constraints are competitive with, and in certain cases surpass, existing bounds obtained from accelerator, reactor, atmospheric, and astrophysical neutrino experiments. Notably, we find that recent advancements in DD experiments have begun to probe the parameter space approaching the sensitivity of constraints derived from supernova observations and cosmological considerations.

The sterile neutrino remains one of the most compelling extensions to the SM, particularly as it may provide insight into the origin of neutrino mass. Our study demonstrates the capability of current-generation DD experiments, using solar neutrinos via elastic neutrino–electron scattering, to explore this phenomenological scenario in a complementary and competitive manner.
Looking ahead, future improvements in detector exposure, background suppression, and lower recoil energy thresholds are anticipated to further enhance the sensitivity to neutrino electromagnetic properties. Such developments will enable the next generation of experiments to probe transition magnetic moments at levels previously accessible only via other astrophysical or cosmological observations, offering valuable new tests of BSM physics.

\section*{Acknowledgments}
This work was supported by the Scientific and Technological Research Council of Türkiye (TUBITAK) under the project no: 124F416.
\bibliographystyle{unsrt}

\end{document}